\documentclass[a4paper,fleqn,usenatbib]{mnras}

\usepackage[T1]{fontenc}
\usepackage{ae,aecompl}

\usepackage{graphicx}	
\usepackage{amsmath}	
\usepackage{amssymb}	
\usepackage{footnote}
\usepackage{lscape}
\usepackage{float}            
\usepackage[flushleft]{threeparttable}  

\newcommand{\mso}{$\mathcal{M}_{\odot}$}    
\newcommand{\ngc}{$N_\textrm{GC}$}  
\newcommand{\sn} {$S_{\textrm N }$}

\newcommand{\Rad}{$R_{\textrm{e}}$}

\title[Luminosity functions of globular clusters]{Luminosity functions of globular clusters in five nearby spiral galaxies using HST/ACS images}

\author[Lomel\'i-N\'u\~nez et al.]{
Luis Lomel\'i-N\'u\~nez\thanks{E-mail: luislomeli@inaoep.mx (LLN)},
Y.D. Mayya, L.H. Rodr\'iguez-Merino, P.A. Ovando, \\
\newauthor 
and D. Rosa-Gonz\'alez 
\\
Instituto Nacional de Astrof\'isica \'Optica y Electr\'onica, Luis Enrique Erro 1, Tonantzintla 72840, Puebla, Mexico\\
}

\date{Accepted XXX. Received YYY; in original form ZZZ}

\pubyear{2021}

\begin{document}
\label{firstpage}
\pagerange{\pageref{firstpage}--\pageref{lastpage}}
\maketitle

\begin{abstract}
We here present the luminosity function (LF) of globular clusters (GCs) in five nearby spiral galaxies using the samples of GC candidates selected in Hubble Space Telescope mosaic images in $F435W$, $F555W$ and $F814W$ filters. Our search, which surpasses the fractional area covered by all previous searches in these galaxies, has resulted in the detection of 158 GC candidates in M81, 1123 in M101, 226 in NGC~4258, 293 in M51 and 173 in NGC~628. The LFs constructed from this dataset, after correcting for relatively small contamination from reddened young clusters, are log-normal in nature, which was hitherto established only for the Milky Way (MW) and Andromeda among spiral galaxies. The magnitude at the turn-over (TO) corresponds to $M_{\rm V}{_0}$(TO)=$-$7.41$\pm$0.14 in four of the galaxies with Hubble types Sc or earlier, in excellent agreement with $M_V(\mathrm{TO})=-7.40\pm0.10$ for the MW. The TO magnitude is equivalent to a mass of $\sim3\times10^5$~M$_\odot$ for an old, metal-poor population. $M_{\rm V}{_0}$(TO) is fainter by $\sim$1.16~magnitude for the fifth galaxy, M\,101, which is of Hubble type Scd. The TO dependence on Hubble type implies that the GCs in early-type spirals are classical GCs, which have a universal TO, whereas the GC population in late-type galaxies is dominated by old disk clusters, which are in general less massive. The radial density distribution of GCs in our sample galaxies follows the S\'ersic function with exponential power-law indices, and effective radii of 4.0--9.5~kpc. GCs in the sample galaxies have a mean specific frequency of 1.10$\pm$0.24, after correcting for magnitude and radial incompleteness factors.
\end{abstract}


\section{Introduction}

Globular clusters (GCs) are among the oldest objects in the universe.
Their relatively high luminosities ($M_V=-5$ to $-10$) and compact sizes
(half-light radius of a few parsecs) allow them to be readily detectable in 
nearby external galaxies \citep{Harris:1996}. Their low metallicities 
($\log$[Fe/H]$\lesssim-$0.5) and enhancements of $\alpha$ elements ($\log$[O/Fe]$\geq$ 0.20) 
suggest that they are formed at very early epochs of 
galaxy formation \citep{Binney:1998}. 
The GC formation requires highly efficient star formation, usually associated with
intense starburst activity \citep{Bastian:2008}. Star formation in the spheroids 
(early-type galaxies, spiral bulges and halos) represents one such activity in the early universe.
Interactions and mergers between galaxies provided the next epochs of 
star formation efficient enough to form massive star clusters such as GCs
\citep{Whitmore:1995:antena}.
These intermediate-age clusters are similar in size and mass as the young massive clusters, also known as Super Star Clusters (SSCs), seen in presently active 
starburst regions \citep{OConnell:1995}.
In the present work, we refer to as SSCs all those relatively younger clusters associated to disks 
of galaxies and reserve the word GCs to describe old clusters associated to spheroids. 
Because of their early formation, the properties of GC systems in galaxies provide important constraints on
models of galaxy formation and evolution \citep{Ashman:1998}.

A variety of GC system properties that are potentially relevant to cosmological theories of
galaxy formation have been identified. These include,
colour distribution \citep{Larsen:2001, West:2004}, 
luminosity function \citep{Reed:1994, Whitmore:1995}, 
radial density distribution \citep{Kartha:2014, Bassino:2006}, 
specific frequency as a function of galaxy type \citep{Harris:1981, Peng:2008}, 
total number of GCs as a function of supermassive black hole masses 
\citep{Burkert:2010, Harris:2011, Harris:2014}, and the nature of their size 
distribution \citep{Kundu:1998, Larsen:2001, Webb:2012}. 
These properties have been exhaustively reviewed in \citet{Brodie:2006}.

Elliptical galaxies have been the most commonly used targets for the study of GC systems.
This is principally because of the relative ease with which GCs can be identified 
when superposed on a smooth light distribution in these galaxies, as compared to 
the inhomogeneities inherent to the disks of galaxies. In addition, the identification
procedure of GCs in disk galaxies has to take into account possible 
contamination from reddened young SSCs and intermediate-old (age$\sim$1--10~Gyr) SSCs. 
Correction of GC magnitudes and colours for the effects of dust in the interstellar 
medium also becomes more important in spiral galaxies, than in elliptical galaxies.

The most important characteristic of GC systems is the 
existence of bimodality in their colour distribution 
\citep{Zepf:1993, Gebhardt:1999, Larsen:2001}. 
This bimodality is understood to be due to an underlying bimodal distribution 
of GC metallicities \citep{Brodie:2006}. Elliptical galaxies with well-determined metallicities confirm
the existence of such a bimodal distribution in metallicities \citep[e.g.,][]{Cohen:1998, Usher:2012}.
This has led to the division of GC systems into two sub-populations: red and 
blue, corresponding to metal-rich and metal-poor populations, respectively. 
These two sub-populations show differences in other properties,
such as size and spatial distribution \citep{Kundu:1998, LarsenBro:2000, Webb:2012}, suggesting possibly two independent formation channels of GC systems.

The GC luminosity function (GCLF) is another property that is well established
in elliptical galaxies. GCLFs are described by 
log-normal distributions with $M_{\textrm 0, V}=-7.40\pm0.25$ and 
$\sigma=1.40\pm0.06$ \citep[e.g.,][]{Reed:1994, Whitmore:1995}. 
The GCLF is often suggested to be a universal function 
\citep[e.g.,][]{Hanes2:1977, Richtler:2003} in elliptical galaxies,
and have been used as a standard candle for the determination of distances to their host galaxies \citep[e.g.][]{Richtler:2003}. 
Among spiral galaxies, the log-normal nature of LFs have been firmly
established only in the Milky Way (e.g., \citealt{Harris:1996}; \citealt{Bica:2006}) 
and Andromeda (e.g., \citealt{Peacock:2010}), the only two spiral galaxies where
GC system properties have been reasonably well-characterized, with values of
$M_{\textrm 0,V}=-7.33$, $\sigma=1.23$ \citep{Secker:1992, Reed:1994} 
and $M_{\textrm 0,V}=-7.6\pm 0.15$, $\sigma=0.82$ \citep{Secker:1992, Reed:1994}, 
respectively. Within the errors of measurements, these values are
in agreement with the TO found in elliptical galaxies \citep{Ashman:1994}.
Using the catalog of \citet{Harris:1996} with 143 GCs, \citet{Jordan:2007}
found $M_{\textrm 0,V}=-7.40\pm0.10$, $\sigma=1.15\pm0.10$ for MW.
\citet{Fall:2001} reproduced the log-normal form, as well as the TO value of the Galactic 
GCs by evolving dynamically clusters obeying power-law mass functions under the gravitational 
potential of the Milky Way.
The uniformity in the TO values in spite
of GC systems being constituted of two independently formed sub-populations
puts strong constraints on the formation scenarios of metal-rich
and metal-poor sub-populations. 

\begin{table*}
        \setlength\tabcolsep{2.3pt}
	\centering 
        \begin{scriptsize}
	\caption{Sample of spiral galaxies}
	\begin{tabular}{l l r r c r r r r r r c l r l c c c c} 
\hline 
Name      & Hubble    &  RA      &DEC       & A$_{V}$ & R$_{25}$ & a$_{\rm max}$     &  d    &$m-M$             &  Distance     & Source  &  M$_{V_{0}}$ & $(B-V)_{0}$ & Scale  & Source  & Proposal   & Number\\
          & Type      &  (J2000) & (J2000)  & (mag)   & (\arcmin)& (\arcmin)     & (Mpc) & (mag)                  & method       &          &             &             & (pc)   & SSCs    & ID         & of fields\\
(1)       & (2)       &  (3)     & (4)      & (5)     & (6)      & (7)           &  (8)  &(9)              &  (10)         & (11)    &   (12)       & (13)        & (14)   & (15)    & (16)       & (17)  \\
\hline
\hline                            
M81       & Sab       &  09:55:33.1 & 69:03:55  & 0.220   & 13.45    & 7.74      & 3.61  & 27.79$\pm$0.06   &  Cepheids     & 1        &  $-$21.10   & 0.91        &  0.87  & 1,2     & 11570      &  29      \\
M101      & Scd       &  14:03:12.5 & 54:20:56  & 0.023   & 14.42    & 7.49      & 6.95  & 29.21$\pm$0.06   &  Cepheids     & 2        &  $-$21.37   & 0.44        &  1.68  & 4,6,7   & 9490,9492  & 12       \\
NGC4258 & Sbc       &  12:18:57.5 & 47:18:14  & 0.045     &  9.31    & 8.60      & 7.576 & 29.397$\pm$0.032 &  MASER        & 3        &  $-$21.03   & 0.67        &  1.84  & 3       & 1157       & 17       \\
M51       & Sbc       &  13:29:56.2 & 47:13:50  & 0.095   &  5.61    & 3.40      & 8.43  & 29.67$\pm$0.02   &  SNII         & 4        &  $-$21.40   & 0.57        &  2.04  & 4,5     & 10452      & 6(4)     \\
NGC628  & Sc        &  01:36:41.7 & 15:47:01  & 0.192     &  5.23    & 5.1       & 9.77  & 29.95$\pm$0.04   &  SNII         & 5        &  $-$20.75   & 0.50        &  2.36  &  4,8    & 10402      & 3        \\
\hline
	\end{tabular}
            \begin{tablenotes}
                \begin{small}
                \item {\it Notes:} (1) Galaxy Name, (2) Hubble morphological type from RC3 \citep{RC3}, (3,4) Right ascension, Declination, in J2000, (5): Galactic extinction from \citet{Schlafly:2011}, (6) $R_{25}$ from RC3 \citep{RC3}, (7) $a_{\rm max}$ is the angular size along the semi-major axis covered by observations, (8) Distance used in this work, (9) Distance modulus, (10) Distance estimation method, (11) Source of distances: 1.- \citet{Tully:2013}; 2.- \citet{Riess:2016}. 3.- \citet{Reid:2019}; 4.- \citet{Rodriguez:2014}; 5.- \citet{Olivares:2010}, (12,13) Absolute magnitude in $V$ band and $(B-V)_{0}$ colour from RC3 \citep{RC3}, (14) HST/ACS pixel scale in pc\,pixel$^{-1}$, (15) References to previous studies of stellar clusters: 1.- \citet{Mayra:2010}; 2.- \citet{Nantais:2010}; 3.- \citet{Lomeli:2017}; 4.- \citet{Whitmore:2014}; 5.- \citet{Hwang:2008}; 6 \citet{Barmby:2006}; 7 \citet{Simanton:2015}; 8 \citet{Adamo:2017}, (16) HST proposal number, (17) Number of ACS fields of each galaxy used in this work.
                \end{small}
           \end{tablenotes}
        \label{tabla:muestra}
        \end{scriptsize}
\end{table*}

The total number of GCs in galaxies (\ngc) is another property that strongly
constraints the formation scenarios of galaxies. 
\citet{Hanes:1977} found that \ngc\ is proportional 
to the mass of its host galaxy and is independent of its morphology.  
\citet{Harris:1981} defined the specific frequency, \sn, defined as the number 
of GCs per absolute magnitude in the $V$-band, normalized to $M_{V}=-15$~mag: 
\sn\ $= {N}_{\text G\text C}\times10^{0.4({M}_{V}+15)}$. They found that \sn\ of 
a galaxy is approximately proportional to the total luminosity of the spheroidal 
component in the galaxy. Recent studies have shown that elliptical galaxies 
have a higher \sn\ as compared to that in spiral galaxies 
(e.g., \citealt{Peng:2008}; \citealt{Georgiev:2010}). 

Classical models of formation of elliptical galaxies either by merging of disk galaxies 
or by the multi-phase dissipational collapse have serious short comings to explain all 
the above-discussed properties of GC systems. For example, the scenario of
the formation of ellipticals by mergers of spiral galaxies predicts similar \sn\ 
values in ellipticals as compared to spirals. These models require the metal-poor 
population forming earlier than the metal-rich clusters. On the other hand,
under the hierarchical scenario of galaxy formation, metal-rich GC systems formed
in-situ in the parent galaxy, which are defined as the highest peaks in density 
fluctuations, whereas the metal-poor GC systems formed in low-mass halos and got accreted 
into host galaxy \citep{Cote:1998}. The predictions from these latter models depend 
critically on the assumed GC system properties in spiral and dwarf galaxies, which are
yet to be well established beyond the Milky Way and Andromeda.

For firmly establishing GC system properties in spiral galaxies, the 
GC sample should cover a substantial area of the galaxy, reaching 
limiting magnitudes fainter than the expected TO magnitudes. In addition, 
it is desirable to have observations in one of the filters blueward of the
Balmer jump ($\sim$3650~\AA) at least in some fields in order to estimate 
the contamination of the GC sample by reddened SSCs.
The first criterion requires analysis of multiple pointing Hubble Space Telescope
(HST) images or ground based images taken with wide-field cameras, whereas the
second one requires images deep enough to register $V\sim$23~mag at distances less than 10~Mpc. 
The third criterion requires the availability of images in
the $U$-band covering at least for some part of the target galaxies.

The existing works on spiral galaxies do not fulfil all the above three criteria.
For example, \citealt{Chandar5galaxias:2004} carried out a study of GCs in five spiral galaxies 
and \citet{Goudfrooij:2003} of seven edge-on disk galaxies, both using HST/WFPC2 fields, but 
covered relatively small areas in each galaxy.
Study by \citet{Young:2012} of two edge-on spirals covered spatially
their entire optical extents, but had a limiting magnitude slightly brighter
than the expected TO. Other works on spiral galaxies include
\citealt{Mayra:2010} and \citealt{Nantaisfoto:2010} (M81; Sab),
\citealt{Hwang:2008} (M51; Sbc), 
\citealt{Barmby:2006}; \citealt{Simanton:2015} (M101; Scd), 
\citealt{Cantiello:2017} (NGC~253; Sc) and \citealt{Lomeli:2017} (NGC~4258 Sbc).
None of these works fulfilled all the three criteria.

\begin{figure*}
    \includegraphics[width=\columnwidth]{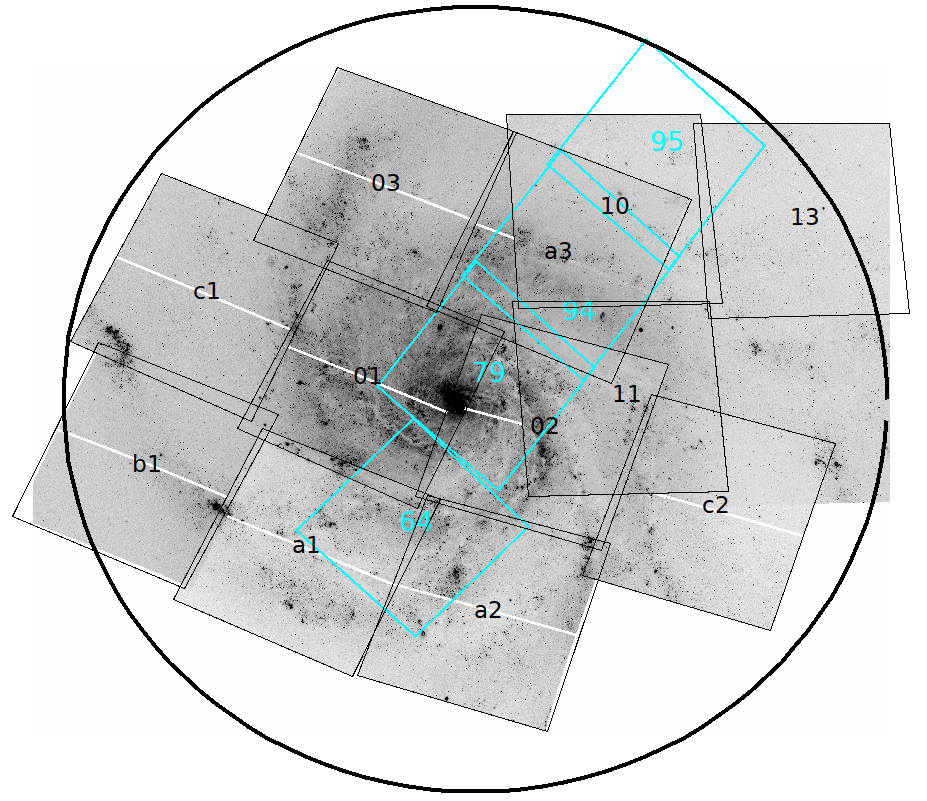}
    \includegraphics[width=\columnwidth]{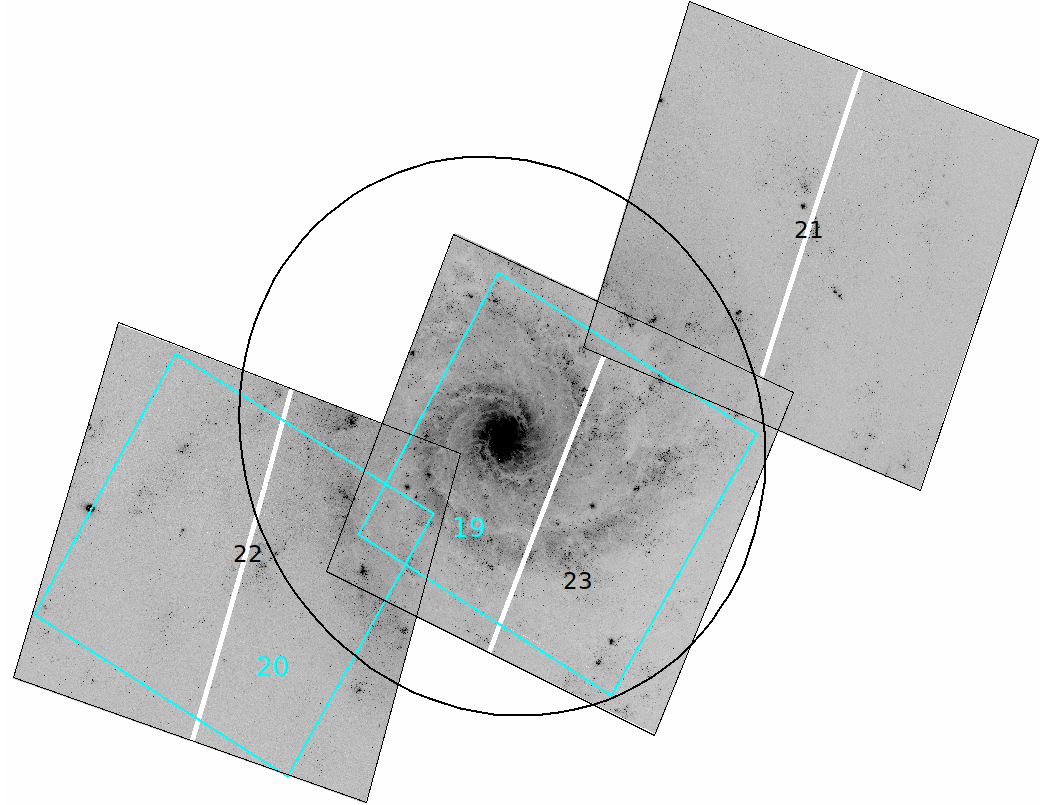}
    \includegraphics[height=10cm,width=\columnwidth]{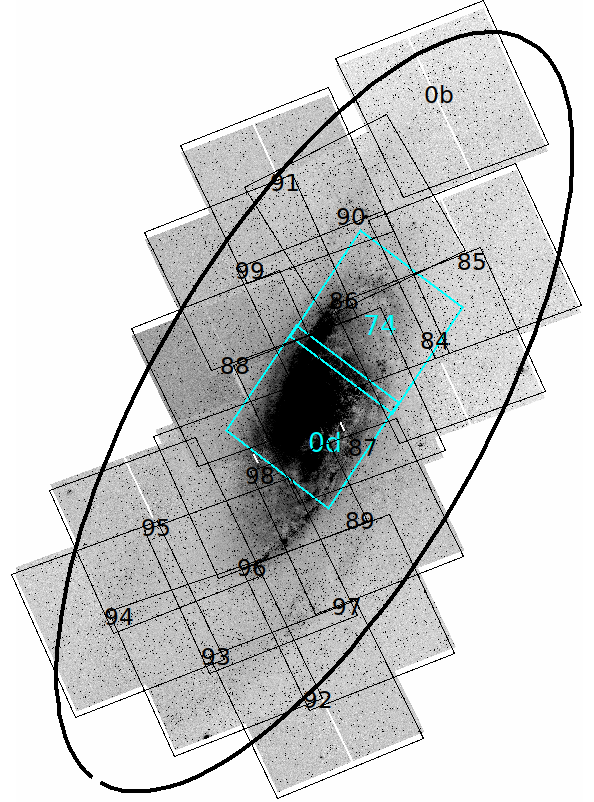}
    \includegraphics[height=10cm,width=\columnwidth]{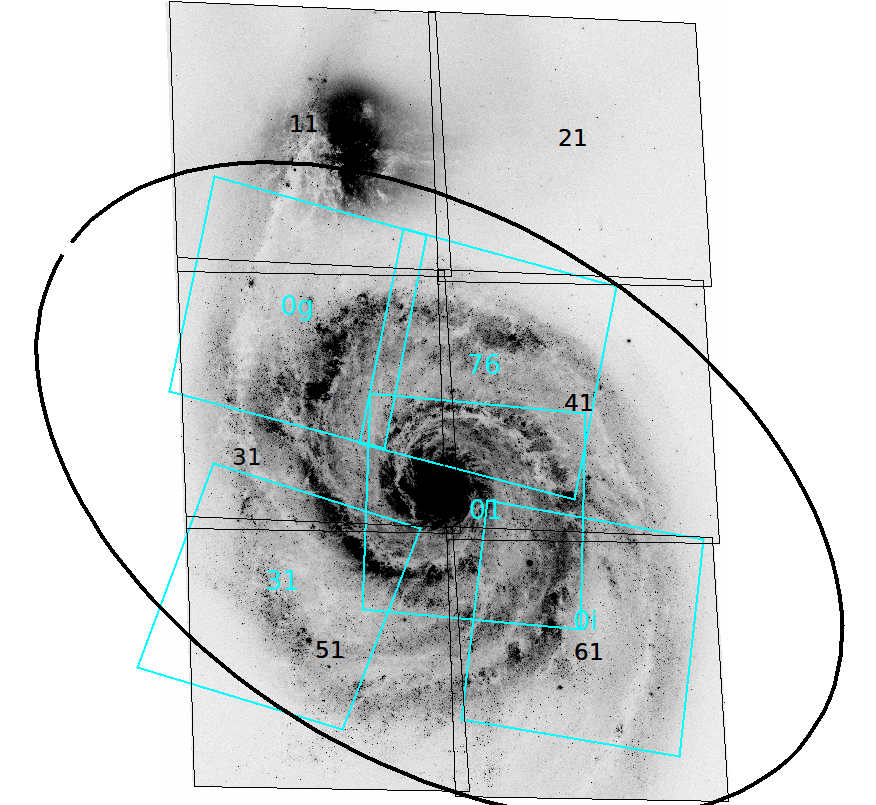}
    \caption{Sample of study. Galaxies are shown in $F435W$ filter, with observation 
footprints superimposed. Cyan-coloured rectangles show the $F336W$ filter footprints. 
The number in each footprint indicates the visit number within proposal. 
All galaxies are aligned such that north is up and east to the left. 
{\it Black ellipse} represent $R_{25}$ for NGC4258 and M51, 
and 0.5$R_{25}$ for M101 and NGC 628. 
See Figure~1 in \citet{Mayra:2010} for visualization of footprints in M81.}
    \label{fig:muestra_galaxias}
\end{figure*}

At present, multi-pointing HST data covering a good fraction of the galaxy 
angular size (see Figure \ref{fig:muestra_galaxias}) and observed in multiple filters
exist for a number of nearby spiral galaxies. 
Some of these galaxies are also observed in the $F336W$ filter.
A careful analysis of these images would be able to provide catalogues
of GCs that are free from contaminating stellar and non-stellar objects, and 
with a good spatial coverage. 
In this paper, we analyse 
five giant spiral galaxies that fulfill all the criteria mentioned above, and are
nearer than 10~Mpc. 

In \S2, we describe the sample of galaxies and the data used. In \S3, 
we explain the detection and our criteria for selecting sample of GCs 
in the selected galaxies. Results of the completeness tests are presented 
in \S4. Analysis and discussion of the properties of our GC samples is presented 
in \S5. In \S6, we give concluding remarks.

\section{Sample of spiral galaxies and data}

With the aim of comparing the properties of GC populations in spiral galaxies with those in 
elliptical galaxies, we searched for nearby giant spiral galaxies with multiple 
pointing HST Advanced Camera for Surveys (ACS) images in at least three optical 
broadband filters. Galactic GCs typically have 
half-light radius $r_{\rm h}\lesssim10$~pc \citep{Harris:1996}. At the spatial 
resolution of $\sim$0.1\,arcsec (image scale=0.05\,arcsec\,pixel$^{-1}$) 
offered by the ACS, majority of the GCs are
 more extended than the Point Spread Function (PSF) up to $\sim$10~Mpc distances.
 Beyond this distance, GC selection would be heavily affected by incompleteness.
 Keeping this in mind, we looked for HST images of giant spiral galaxies at 
distances $<$10~Mpc. We analysed such data for five galaxies. In Table \ref{tabla:muestra}, 
we list these galaxies along with their basic properties and source of images. 
All these galaxies have multiple pointing HST/ACS images in 
$F435W$ ($B$), $F555W$ ($V$), and $F814W$ ($I$) bands, covering a 
good fraction of their optical extent. In addition, all have at least one pointing in 
the $F336W$ ($U$) filter, which allows an estimation of the contamination of 
our GC catalogues by reddened SSCs.

The Hubble Legacy Archive\footnote{\url{https://hla.stsci.edu/hlaview.html}} (HLA)
provides images and photometric catalogues obtained with 
DAOPHOT\footnote{\url{http://www.star.bris.ac.uk/~mbt/daophot/}} \citep{DAOPHOT:1987} 
and SExtractor\footnote{\url{https://www.astromatic.net/software/sextractor}} \citep{Sextractor:1996}. 
These catalogues have been used by \citet{Whitmore:2014} to study the luminosity 
function of star clusters in selected fields for a sample of 20 spiral and 
irregular galaxies. As a first step, we used these catalogues to select a sample 
of GCs. However, we noticed that the adjacent images
had vastly different limiting magnitudes in some galaxies. A visual inspection of
catalogued sources on the images suggested that the background values used
in some images were inappropriate (see Appendix \ref{apen:comparasion}).
Since we are looking for a complete
sample of GCs up to a given limiting magnitude, we decided to make our own 
catalogues on the downloaded images.

In Figure~\ref{fig:muestra_galaxias}, we show the images of galaxies in $F435W$ band, 
with footprints superposed. In Tables~\ref{tabla:apuntados} and \ref{tabla:apuntados_f336w}, 
we give logs of observations for four of our galaxies in the optical and $F336W$ filters, 
respectively. Different columns tabulate
pointing IDs, exposure times, zeropoints\footnote{The zeropoints were obtained from ACS Zeropoints Calculator: 
\url{https://acszeropoints.stsci.edu/}.} 
($c0$) in each filter. We also give the number of stars used for astrometry and 
RMS error in the coordinates. The fifth galaxy, M81, has been the subject of study 
previously by our group \citep{Mayra:2010}, and hence we directly use the GC 
catalog from that study.

\begin{table}
\setlength\tabcolsep{1.6pt}
\begin{center}
\begin{small}
\caption{Log of HST/ACS optical observations of our sample galaxies$\dagger$.}
\begin{tabular}{l c c c c c c c c c c }     
\hline
 & \multicolumn{2}{|c|}{$F435W$}  &  \multicolumn{2}{|c|}{$F555W$}  & \multicolumn{2}{|c|}{$F814W$} & \multicolumn{2}{|c|}{Astrometric} \\ 
          ID   & $T_{\text{exp}}$ & $c0$  & $T_{\text{exp}}$  & $c0$  &  $T_{\text{exp}}$   & $c0$    & $N_{\rm stars}$ & RMS\\
               & (s)   &                             &   (s)  &                             &    (s)  &      &       & (\,arcsec)\\
\hline
\hline    
\multicolumn{9}{|c|}{M101} \\
\hline
             01        & 900   & 25.792  &   720 & 25.736   &   720  & 25.531 & 93 & 0.0259 \\             
             02        & 900   & 25.792  &   720 & 25.736   &   720  & 25.531 & 55 & 0.163  \\             
             03        & 900   & 25.792  &   720 & 25.736   &   720  & 25.531 & 38 & 0.0385 \\             
             a1        & 900   & 25.792  &   720 & 25.736   &   720  & 25.531 & 41 & 0.0319 \\            
             a2        & 900   & 25.792  &   720 & 25.736   &   720  & 25.531 & 37 & 0.278  \\             
             a3        & 900   & 25.792  &   720 & 25.736   &   720  & 25.531 & 11 & 0.0766 \\             
             b1        & 900   & 25.792  &   720 & 25.736   &   720  & 25.531 & 10 & 0.187  \\             
             c1        & 900   & 25.792  &   720 & 25.736   &   720  & 25.531 & 18 & 0.0925 \\             
             c2        & 900   & 25.792  &   720 & 25.736   &   720  & 25.531 & 21 & 0.0175 \\             
             10        & 1080  & 25.792  &  1080 & 25.736   &   1080 & 25.531 & 13 & 0.0988 \\             
             11        & 1080  & 25.792  &  1080 & 25.736   &   1080 & 25.531 & 29 & 0.0456 \\             
             13        & 1080  & 25.792  &  1080 & 25.736   &   1080 & 25.531 & 11 & 0.0862 \\             
\hline
\multicolumn{9}{|c|}{NGC 4258} \\
\hline                                                                                                     
             0b        & 360   & 25.767  &  975  & 25.717   &   375 & 25.520  & 6  & 0.0134   \\           
             84        & 360   & 25.767  &  975  & 25.717   &   375 & 25.520  & 8  & 0.0264  \\ 
             85        & 360   & 25.767  &  975  & 25.717   &   375 & 25.520  & 6  & 0.0058 \\
             86        & 360   & 25.767  &  975  & 25.717   &   375 & 25.520  & 27 & 0.0049   \\ 
             87        & 360   & 25.767  &  975  & 25.717   &   375 & 25.520  & 35 & 0.0049   \\
             88        & 360   & 25.767  &  975  & 25.717   &   375 & 25.520  & 26 & 0.0051  \\
             89        & 360   & 25.768  &  975  & 25.717   &   375 & 25.521  & 12 & 0.0023  \\
             90        & 360   & 25.767  &  975  & 25.717   &   375 & 25.520  & 13 & 0.0090  \\
             91        & 360   & 25.767  &  975  & 25.717   &   375 & 25.520  &  5 & 0.039    \\
             92        & 360   & 25.768  &  975  & 25.717   &   375 & 25.521  &  8 & 0.025    \\
             93        & 360   & 25.768  &  975  & 25.717   &   375 & 25.521  &  9 & 0.0273   \\
             94        & 360   & 25.768  &  975  & 25.717   &   375 & 25.521  &  9 & 0.03     \\ 
             95        & 360   & 25.768  &  975  & 25.717   &   375 & 25.521  &  8 & 0.0145   \\
             96        & 360   & 25.768  &  975  & 25.717   &   375 & 25.521  & 20 & 0.0157   \\
             97        & 360   & 25.768  &  975  & 25.717   &   375 & 25.521  &  8 & 0.0054  \\
             98        & 360   & 25.768  &  975  & 25.717   &   375 & 25.521  & 30 & 0.0054  \\
             99        & 360   & 25.767  &  975  & 25.717   &   375 & 25.520  & 14 & 0.0128   \\
\hline
\multicolumn{9}{|c|}{M51} \\
\hline  
            1-6  & 680$\times$4  & 25.888  &  340$\times$4 & 25.715  &   340$\times$4 & 25.471 & 299  & 0.0482 \\
\hline
\multicolumn{9}{|c|}{NGC 628} \\
\hline
            21        &  1200  & 25.789  &  1000 & 25.732  &   900 & 25.528  & 9   & 0.0445 \\
            22        &   800  & 25.788  &  360  & 25.731  &   720 & 25.528  & 15  & 0.047 \\
            23        &  1358  & 25.789  &  858  & 25.732  &   922 & 25.528  & 15  & 0.034 \\
\hline
\end{tabular}
\label{tabla:apuntados} 
\begin{tablenotes}
    $^\dagger$Table~1 of \citet{Mayra:2010} contains observational log for 29 pointing in M81, the fifth galaxy of our sample.
\end{tablenotes}
\end{small} 
\end{center} 
\end{table}

\begin{table}
\setlength\tabcolsep{2.5pt}
\begin{center}
\begin{small}
\caption{Log of HST/WFC3 $F336W$-band observations of our sample galaxies.}
\begin{tabular}{l c c c c c c c c c c c c c c c c c c c c c c c c}     
\hline
ID   & $T_{\text{exp}}$ & $c0$  & $N_{\rm stars}$ & RMS     & Proposal \\ 
     & (s)              &       &                 &(\,arcsec)&  ID      \\ 
(1)  & (2)              & (3)   & (4)             & (5)     &  (6)      \\
\hline
\hline
\multicolumn{6}{|c|}{M101}                                     \\
\hline    
        64  &   2361 & 23.546 & - & - & 13364     \\    
        79  &   2382 & 23.546 & - & - & 13364     \\    
        94  &   2382 & 23.546 & - & - & 13364      \\   
        95  &   2382 & 23.546 & - & - & 13364      \\   
\hline
\multicolumn{6}{|c|}{NGC 4258} \\
\hline                  
0d  &   1062 & 23.546  & 33 &   0.0154 &13364  \\ 
74  &   1062 & 23.546  & 15 &   0.0109 &13364  \\ 
\hline
\multicolumn{6}{|c|}{M51}                       \\
\hline
 01   &  4360 & 23.546  &  163 & 0.0352 &13340  \\ 
 0g   &  2376 & 23.546  &  55  & 0.0191 &13364  \\ 
 0i   &  2361 & 23.546  &  42  & 0.0317 &13364   \\
 76   &  2376 & 23.546  &  64  & 0.0281 &13364   \\
 31   &   780 & 23.546  &  31  & 0.0350 &14149   \\
\hline
\multicolumn{6}{|c|}{NGC 628}               \\
\hline
 19  &  2361 & 23.546 &   15  & 0.0031 & 13364 \\ 
 20  &  1119 & 23.546 &   10  & 0.0321 & 13364 \\ 
\hline
\end{tabular}
\label{tabla:apuntados_f336w} 
\end{small} 
\end{center} 
\end{table}

\section{Source detection and cluster selection}

We used images in $F814W$ band for detecting sources using SExtarctor. 
The critical detection parameters used are: {\sc detect\_minarea}=5 pixel, 
{\sc detect\_thresh}=1.4, {\sc back\_size}=32 pixel and {\sc back\_filtersize}=3 pixel,
where each pixel corresponds to 0.05\,arcsec.
Using these criteria, typically we obtained several tens of thousands 
sources in each frame. Elaborate filtering criteria need to be implemented 
to select GCs from this catalogue.

As discussed in the introduction, making a catalogue of GCs is more challenging 
in spiral galaxies as compared to elliptical galaxies. This is because, unlike 
elliptical galaxies, spiral galaxies show a lot of structures at the scale comparable 
to the size of GCs. These structures lead to a lot of spurious sources in the SExtarctor catalogue.
Filtering based on structural parameters \citep[e.g.][]{Mayya:2008}, colour 
\citep[e.g.][]{Fedotov:2011}, concentration index \citep[e.g.][]{Whitmore:2014}, colour-colour 
diagrams  \citep[e.g.][]{Munoz:2014, Lomeli:2017} or a combination of these, 
are the most commonly used methods. 
In this paper, we used the cluster selection method used in \citet{Mayya:2008} 
and \citet{Mayra:2010}, which consists of using SExtractor-derived 
structural parameters ({\sc fwhm, area, ellipticity}) to define a cluster sample,
and  photometric parameters ({\sc colour}) to separate young clusters from GCs. 
The method is described in detail in Sections \ref{seccion:seleccion} and \ref{seccion:sample}, below.

\subsection{Astrometric correction of HST images}

\begin{figure}
    \includegraphics[height=3cm,width=\columnwidth]{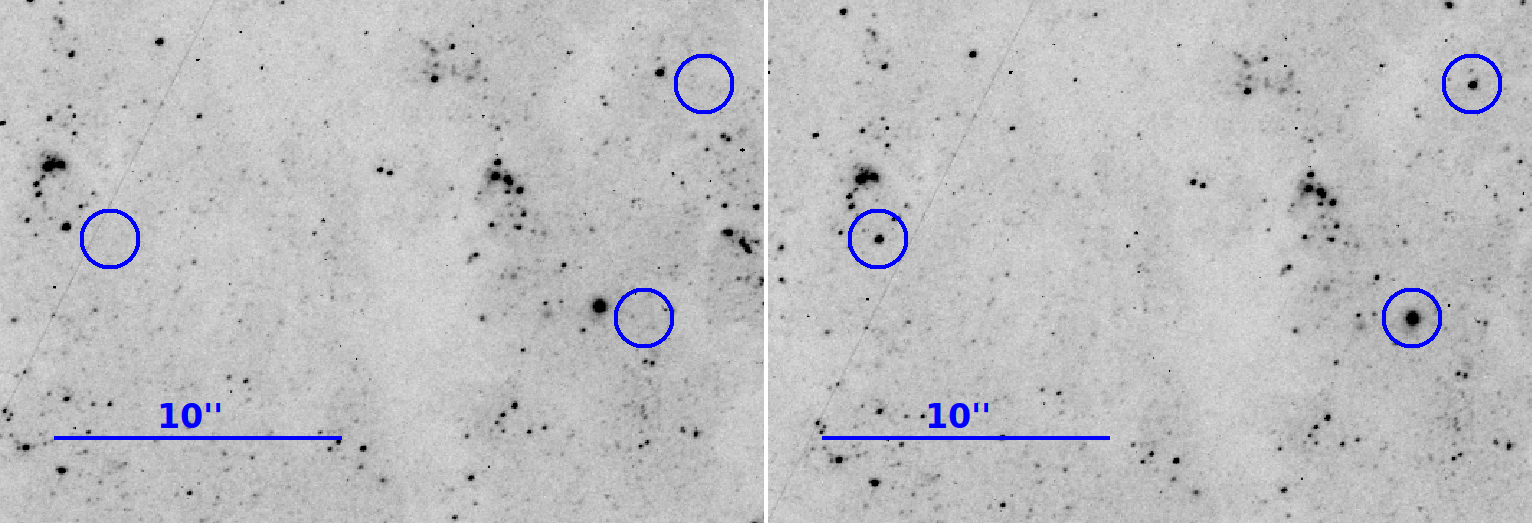} 
    \caption{Illustration of astrometry on the HST image using a section of M101. {\it Left:} 
             astrometric coordinates on the downloaded image from HLA. {\it Right:} image after 
             applying our astrometric corrections based on the GAIA DR2 coordinates of the field stars. 
             Three field stars in the displayed FOV are identified with blue circles of 1\,arcsec radius. 
             North is up and east to the left in these images.  }  
    \label{fig:gaia_correction}
\end{figure}

Before running SExtratctor it was necessary to astrometrize all HLA images. 
We performed this with the help of GAIA stars using Gaia Data Release 2 
(Gaia DR2; \citealt{GAIA:2016}; \citealt{GAIADR2:2018}). 
We used the IRAF task {\it ccmap} with a second order polynomial in coordinates to 
achieve this. Minimum of 10 stars were used in each pointing (5 stars in one pointing 
of NGC 4258), resulting in mean rms astrometric accuracies of $\sim$0.095\,arcsec\ 
(M101), $\sim$0.014\,arcsec\ (NGC\ 4258), $\sim$0.048\,arcsec\ (M51), and $\sim$0.042\,arcsec\ 
(NGC\ 628). We took into account these rms errors to identify and eliminate duplicate sources in the overlap zones between two adjacent frames.
In Figure~\ref{fig:gaia_correction}, we zoom in on an image section in M101 to illustrate
a typical field before and after astrometric correction. 
The left image shows the HST coordinate system for this field, 
whereas the image on the right shows the corrected coordinate system. The circles show 
the coordinates of GAIA (Gaia DR2) stars in this field of view. 

\begin{figure}
	   \includegraphics[width=0.95\columnwidth]{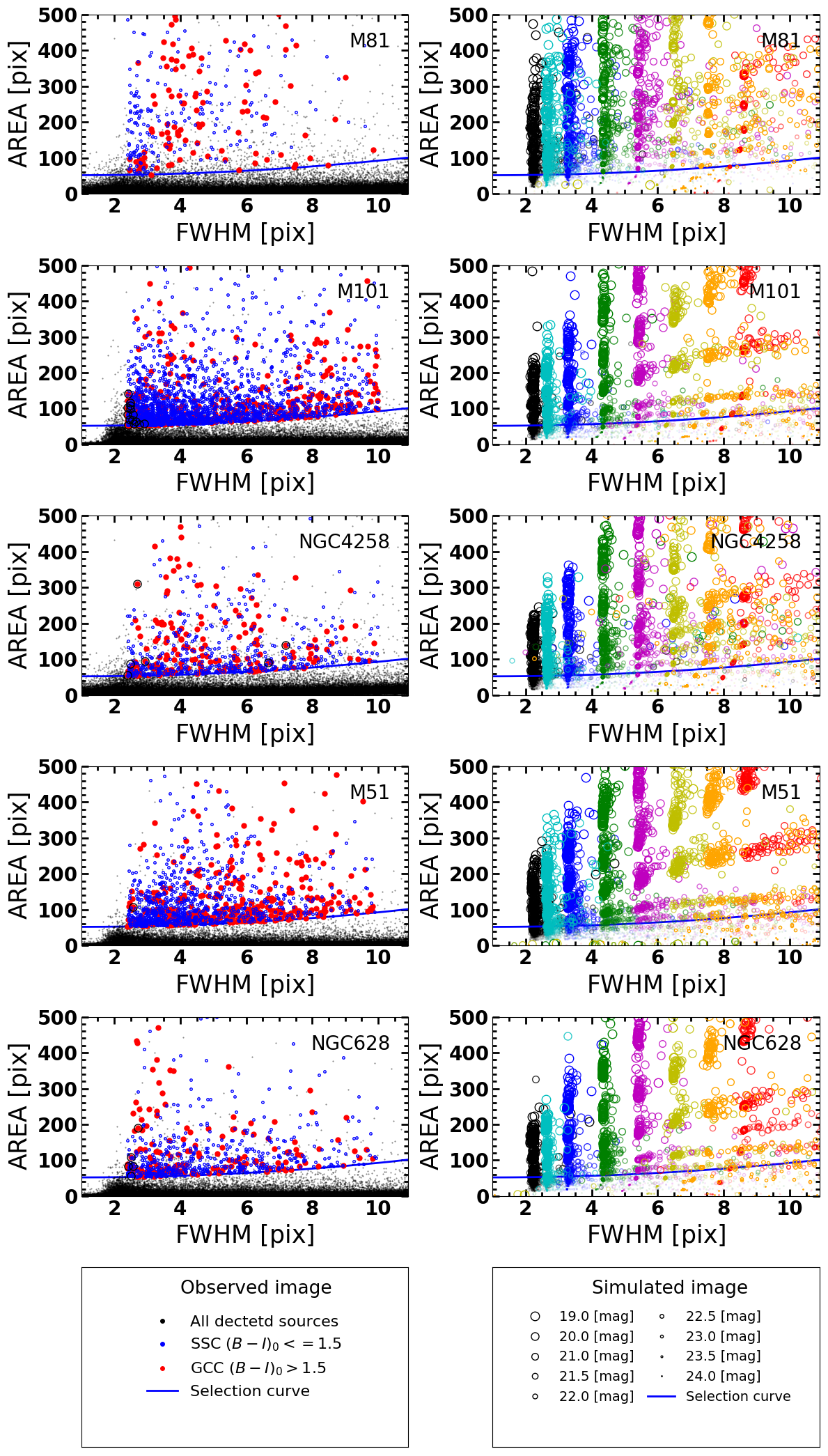}
    \caption{
    {\it Left:}
    Selected GCs (red dots) compared to all SExtractor sources (black dots)
in {\sc area} vs {\sc fwhm} plane for our 5 sample galaxies. 
GCs have $(B-I)_{0}$>1.5~mag, 2.4$<${\sc fwhm}/pix$<$10 and are above the parabola (blue line).
Clusters bluer than this limit are SSCs which are identified
by blue dots. Black dots above the parabola are extended sources with ellipticities$>$0.3,
and hence do not satisfy all cluster selection criteria. 
{\it Right:}
{\sc area} vs {\sc fwhm} for simulated star clusters. The {\sc fwhm} of mock clusters are fixed at
2.0 (black), 2.4 (cyan), 3.0 (blue), 4.0 (green), 5.0 (magenta), 6.0 (yellow), 7.0 (orange) and  8.0 (red) pixels.
For each fixed {\sc fwhm}, 100 mock clusters between 
19-24 magnitudes are generated, which are shown by symbols of different sizes (successively smaller sizes for fainter sources; see the bottom-most right panel
for a guide on the symbol size). Most of the clusters with mag~$>23$ lie
below the parabola defining the minimum {\sc area}-{\sc fwhm} relation, and hence are rejected
by our selection criteria.
}
    \label{fig:area_fwhm}
\end{figure}

\begin{figure}
\begin{center}
         \includegraphics[width=0.75\columnwidth]{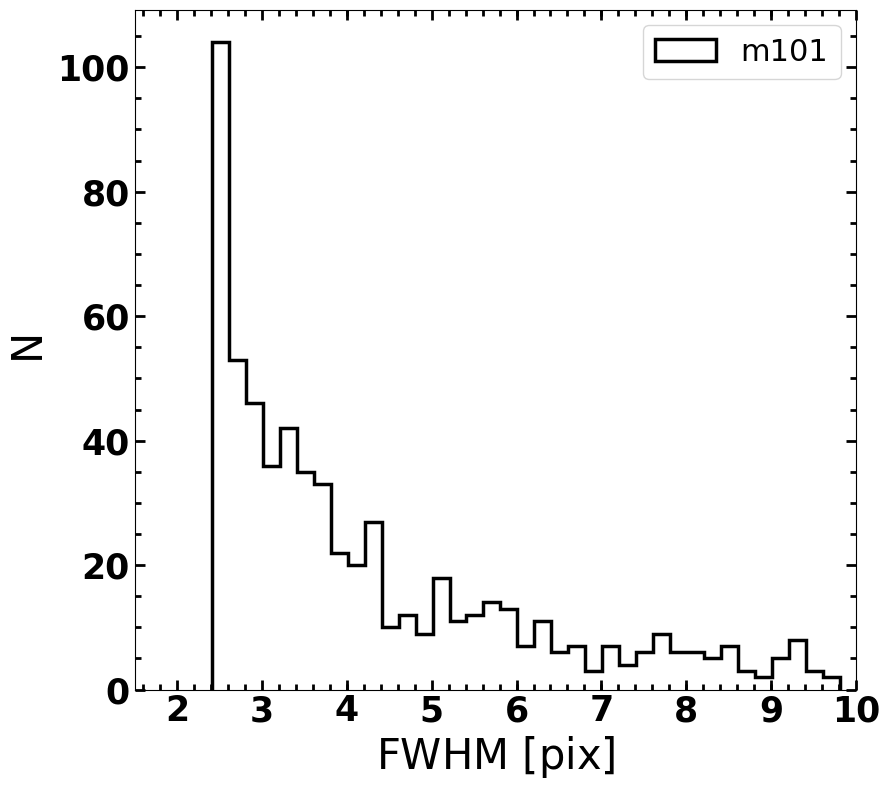}
         \includegraphics[width=0.75\columnwidth]{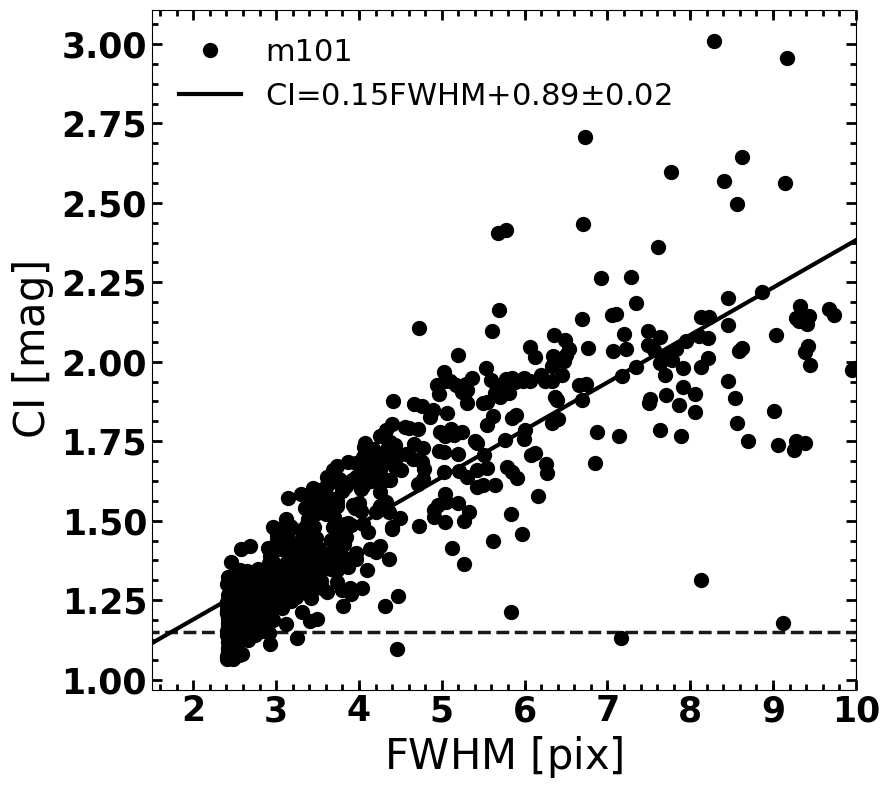}
    \caption{
(Top) Distribution of {\sc fwhm} for the candidate GCs in one of our galaxies (M101).
(Bottom) Comparison of the two commonly used discriminators between clusters and stars, CI vs {\sc fwhm}. 
Data are consistent with a linear fit (solid line whose equation is given in the top left corner) 
above the dividing line between clusters and stars (horizontal dashed line at CI=1.15).
Data correspond to M101, which has 45 GCs below the defining line (see text for details).
}
    \label{fig:ci_vs_fwhm}
\end{center}
\end{figure}

\subsection{Selection criteria for defining a cluster sample}
\label{seccion:seleccion}

In this work, we aim to obtain a sample of GCs with properties similar to those in the Milky Way, which are marginally resolved on the HST/ACS images at the distances of sample galaxies. We considered all objects with {\sc fwhm}$>$2.4~pixel as GC candidates. This cut-off corresponds to 2.1~pc and 5.7~pc, in the nearest (M81) and farthest (NGC~628) galaxies of the sample. Thus, farther a galaxy is, lesser would be the number of compact objects we would detect. GCs do not show a mass-radius (or equivalently luminosity-radius) relationship \citep{Gieles:2010} and hence this bias is not expected to affect the LF of GCs. GCs are roundish objects, and hence their {\sc ellipticity} parameter is expected to be close to zero. SExtractor-measured {\sc ellipticity} even for roundish objects could be as high as 0.3,
as it is measured at the isophote corresponding to the detection threshold on the background subtracted image. 
Another parameter that SExtractor calculates is {\sc area}, with is the number of pixels enclosed by the
isophote where {\sc ellipticity} is measured. Both the {\sc ellipticity} and {\sc area} depend on the threshold used for the detection, and hence a cluster of same magnitude and {\sc fwhm} can have different values of 
{\sc ellipticity} and {\sc area}, depending on the underlying background.

We have carried out Monte Carlo simulations to understand the behaviour of these
parameters for clusters of different {\sc fwhm} and magnitudes, appropriate to the background and crowding 
encountered in each galaxy in the $F814W$ band. In Section \ref{seccion:completes}, we describe in detail these simulations. 
In Figure~\ref{fig:area_fwhm} (right), we show the simulated clusters in {\sc area} vs {\sc fwhm}
diagram for each of our sample galaxies. Simulations were carried out for a range of magnitudes 
between 19 and 24 magnitudes, for fixed values of {\sc fwhm}. As expected, brighter clusters have larger {\sc area}s
at a given {\sc fwhm}, and the {\sc area} increases quadratically with {\sc fwhm} for a cluster of given magnitude.
We show the dependence of {\sc area} with {\sc fwhm} for objects of 23~mag by the blue solid curve, which is
defined by the same equation, ${ \text{AREA}=52.5-1.0\ \text{FWHM} + 0.50\ \text{FWHM}^{2}}$, 
for all our sample galaxies.
On the left panel, we show all the detected sources in {\sc fwhm} vs {\sc area} 
diagram for our five sample galaxies. The parabola defined by the simulations is shown,
which separates the bonafide cluster candidates (that are above the parabola) from
contaminating sources (image blemishes, stellar asterisms, image borders etc.), which dominate the number of detected sources at every {\sc fwhm}.
A hard cut in magnitude or {\sc area} can also eliminate the contaminating objects, but
the use of parabola is the most effective way to eliminate these contaminating sources, 
without missing many genuine clusters.
GCs at the distance of sample galaxies are expected to have {\sc fwhm} close to the
observational lower limit of 2.4~pixels. 
In the top panel of Figure~\ref{fig:ci_vs_fwhm}, we show the observed distribution of 
{\sc fwhm} for our
final GC sample for one of our sample galaxies (M101), which indeed peaks at the 
first bin. However, we include objects up to {\sc fwhm}=$10$~pixels to 
account for possible errors in the measurement of {\sc fwhm} at detection limits.
Simulations suggest that even the brightest clusters ($F814W=19$) do not occupy an {\sc area}$>$500~pixels
as long as the {\sc fwhm}$\sim$5~pixels, and hence we eliminated sources with {\sc area}$>$500~pixel.

The HLA catalogues include a parameter known as concentration index (CI), 
defined as the difference between magnitudes measured in 1 and 3 pixel radius
apertures. Some studies have made use of this parameter to discriminate between point 
and extended sources (e.g., \citealt{Whitmore:2014}, \citealt{Simanton:2015})  
CI $<1.15$ for stars.
In our study, we have used the {\sc fwhm} $=2.4$~pixel (0.12\,arcsec), a direct
discriminator between unresolved (stars) and resolved (extended) sources.
In the bottom panel of Figure \ref{fig:ci_vs_fwhm}, we show CI against the {\sc fwhm} 
for all our clusters for M101. As expected, the two parameters
are correlated. The minimum CI (1.07) for our sample 
clusters is slightly less than 1.15, the value adopted in other studies. 
Forty-five ($\sim$4\%) of our sources wouldn't have been classified as stellar cluster if we 
had adopted the CI criterion. A visual inspection of these borderline
objects suggests they are likely to be clusters, rather than stars. 
We identify these borderline cases (1.07$<$CI$<$1.15) by red dots surrounded by  
black circles in Figure \ref{fig:dcm}. 
In NGC 4258, these include two of the brightest GCs.

\subsection{Aperture photometry}
\label{seccion:aperture_correction}

\subsubsection{Magnitudes, colours and aperture correction}
SExtractor was also used for obtaining photometry in the three bands for all 
the detected sources. The photometry was carried out in 10 apertures with 
radii between 1 and 15 pixel (0.05\,arcsec\ to 0.75\,arcsec), using the zeropoints ($c0$)
tabulated in Table \ref{tabla:apuntados}  for each frame. 
Unlike stars, aperture correction for clusters depend on the cluster size, 
which is parameterized by the {\sc fwhm} in SExtractor. The aperture correction is 
defined as the difference between magnitude at 3 pixel (0.15\,arcsec) radius and the 
magnitude at an infinite aperture ( $\Delta F814W=m_{\rm F814W}$(3 pix)$-m_{\rm F814W}$(tot)). 
Even the most extended clusters (objects with Gaussian {\sc fwhm}=9 pixel) contain more than 
98\% of their total flux within an aperture of 15 pixel radius, because of which we have 
used the magnitude in an aperture of 15~pixel radius as $m_{\rm F814W}$(tot).

\begin{figure}
        \includegraphics[width=0.90\columnwidth]{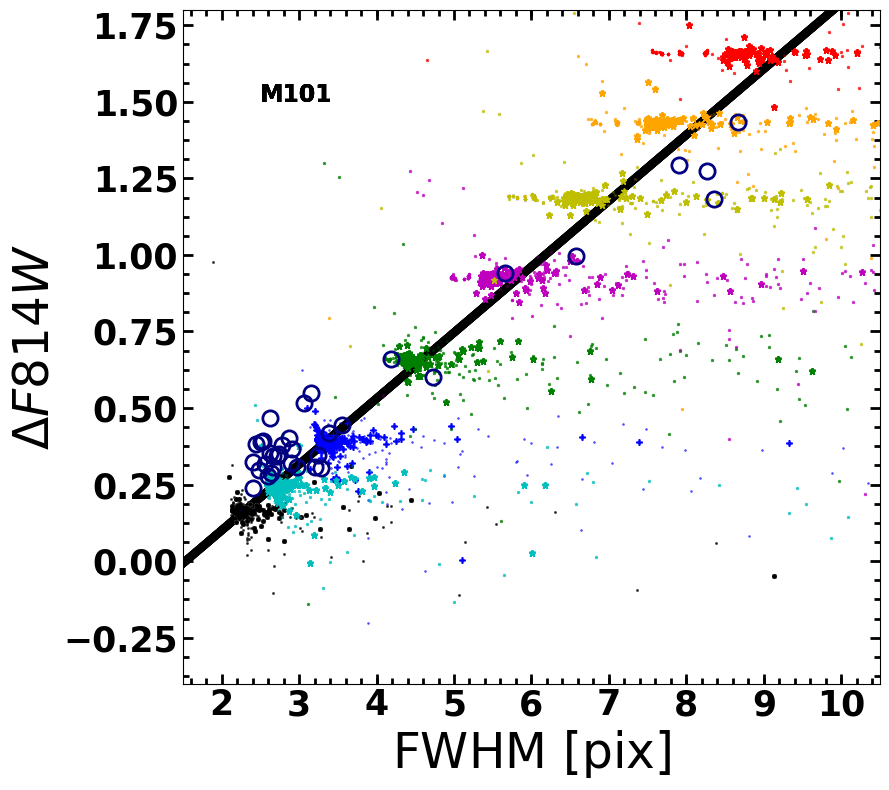}
        \caption{
        Apertures corrections vs FWHM for observed (open circles) and simulated (coloured points) star clusters in M101. The line is the linear fit to the simulated points.}
    \label{fig:nueva_correccion}
\end{figure}

We used the results of our experiments with simulated clusters,
described later in Section~\ref{seccion:completes}, to 
obtain the correction as a function of measured {\sc fwhm} in each galaxy. 
The results are shown in Figure~\ref{fig:nueva_correccion} for M101. 
A group of horizontally distributed points corresponds to the same input {\sc fwhm}. 
The measured {\sc fwhm} for fainter clusters tends to be systematically larger than the input {\sc fwhm}, which is the reason for the horizontal spread. 
The aperture corrections (open circles) obtained for 
35 isolated clusters with good photometry (error $<0.005$~mag in 9~pixel) is also shown
in the Figure. Both the simulated and observed corrections smoothly increase with the 
{\sc fwhm}. For the simulated data, we obtained mean values of 
measured {\sc fwhm} and correction for each input {\sc fwhm} and fitted these mean values
by a straight line which is shown by the solid line. The fitted results (the slope $m$, and 
the abscissa, $b$) for all the sample galaxies are given in Table~\ref{tabla:fit_correction}.

We applied these corrections to the $F814W$ aperture magnitudes of 3~pixel radius 
of all cluster candidates to obtain their total magnitudes. 
Colours ($F435W-F814W$, $F435W-F555W$ and 
$F555W-F814W$), however are obtained by subtracting magnitudes in the 
corresponding filters at 3~pixel radius aperture. This procedure ensures that 
errors on colours are smaller than the aperture corrected magnitudes. 
The errors on colours were calculated by quadratically summing the magnitude errors 
in the two bands forming a colour.

\begin{table}
\begin{center}
\caption{Aperture correction coefficients and typical photometric errors in $(F435W-F555W)_{0}$ colour for the sample galaxies.}
\begin{tabular}{l r r c r r r r}
\hline
Galaxia  & $m$               & $b$   &  $\sigma_{BV}$  \\  
(1)      & (2)             & (3)              & (4)   \\   
\hline
\hline
M81      & 0.193$\pm$0.004 & $-$0.277$\pm$0.026 & 0.10 \\ 
M101     & 0.215$\pm$0.004 & $-$0.330$\pm$0.025 & 0.21 \\ 
NGC~4258 & 0.193$\pm$0.007 & $-$0.279$\pm$0.045 & 0.10 \\ 
M51      & 0.218$\pm$0.004 & $-$0.354$\pm$0.024 & 0.16 \\ 
NGC~628  & 0.225$\pm$0.003 & $-$0.358$\pm$0.022 & 0.19 \\ 
\hline
\end{tabular}
\begin{tablenotes}
 \begin{small}
 \item {{\it Notes:} $\sigma_{BV}$ is the average standard deviation estimated in the $(F435W-F555W)_{0}$ colour 
 in bins of 0.2~mag in $(F435W-F814W)_{0}$ colour for clusters with $(F435W-F814W)_{0}$>1.5. 
 } 
   \end{small}
\end{tablenotes}
\label{tabla:fit_correction} 
\end{center} 
\end{table}

\subsubsection{Photometric error estimation}

\begin{figure}
        \includegraphics[width=0.9\columnwidth]{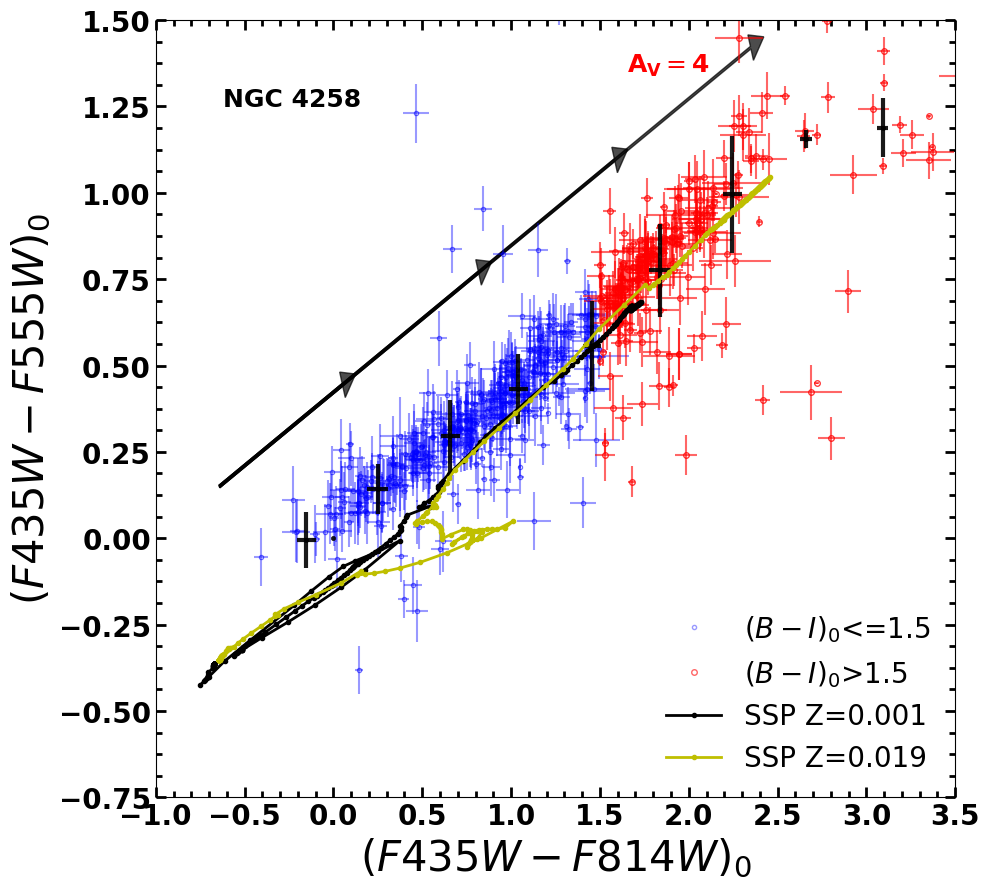}
        \caption{    
Colour-colour diagram showing all SExtractor-detected clusters for one of our sample galaxies 
(NGC~4258). The SSP evolution between 0 to 12~Gyr at two metallicities
and the reddening 
vector for $A_V$=4~mag, with arrow heads placed at 1~mag intervals, are shown. Both the
evolution and the reddening move the points along the same direction, and hence the
spread in either axis around a mean local value (the crosses) is entirely due to 
observational errors in colours. 
}
    \label{fig:error_estimated}
\end{figure}

Formal errors\footnote{\url{https://sextractor.readthedocs.io/en/latest/Photom.html}} 
on photometric measurements are obtained using the formula: {\small 
$ MAGERR=1.086\times(FLUXERR/FLUX)$}, where {\small $FLUXERR=\sqrt{\displaystyle\sum_{i\in A}
(\sigma_{i}^{2}+\frac{p_{i}}{g_{i}}  )} $}, with $A$ representing the set of pixels defining 
the photometric aperture, $\sigma_{i}$, the standard deviation of noise (in ADU) estimated 
from the local background, $p_{i}$ the background-subtracted image pixel value, and $g_{i}$ 
the effective detector gain in e-/ADU at pixel $i$.
However, small-scale variations in the disk background gives rise to additional
errors while carrying out photometry of non-stellar objects such as GCs, 
that are usually larger than the formal errors. We use the colour-colour diagram
$(F435W-F555W)_{0}$ vs $(F435W-F814W)_{0}$ 
to estimate the real errors on our colours. In Figure \ref{fig:error_estimated}, 
we illustrate the method adopted for this. 
In this plot, we show the colours for star clusters in NGC\,4258, with the 
SSPs with Z=0.001 and 0.019 metallicities from \citet{Bruzual:2003}. The reddening vector for 
$A_V$=4~mag is also shown. The reddening vector and the evolutionary trajectory are
parallel in this colour-colour diagram, implying the spread in the $(F435W-F555W)_{0}$ colour 
for a fixed $(F435W-F814W)_{0}$ colour (and vice versa) is entirely due to observational errors. 
We calculated the dispersions in colour for each axis for fixed bins of 0.2~mag width 
in the other axis. In the figure, we show these dispersions by crosses places 
at every 0.4~mag. There is a tendency for slightly higher error for the reddest colours.
We take into account this dispersion in each colour as an additional source of error while
comparing observational colours with model colours.

\begin{figure*}
    \includegraphics[width=16cm]{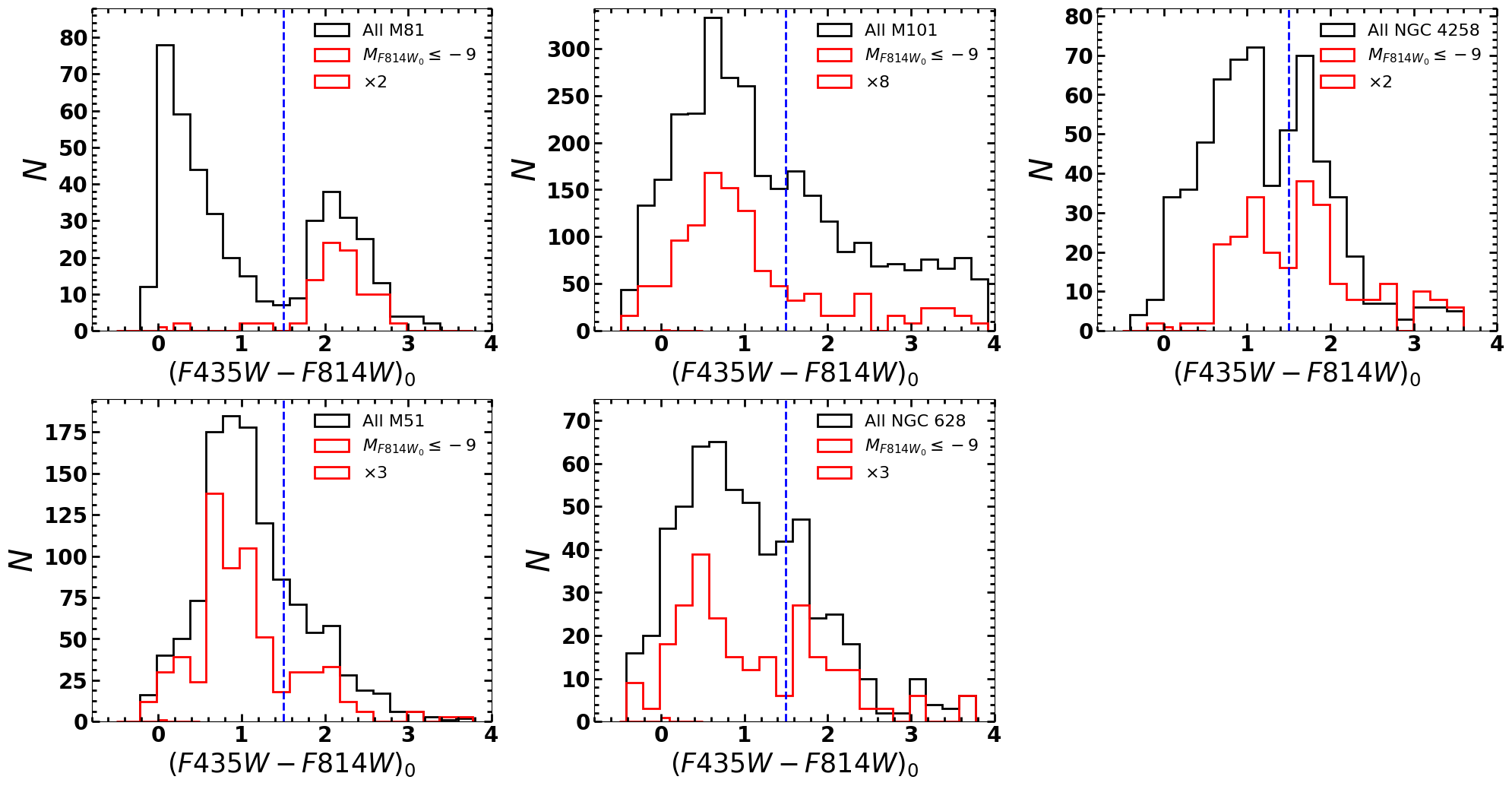}
    \caption{
    {Colour histogram of all (SSC+GC) cluster candidates. The black line shows the colour distribution over the entire range of magnitudes, whereas the red histogram shows the distribution for bright ($M_{I}\leq-$9) clusters. The latter histogram values are multiplied by a factor indicated in each panel. The vertical line at $(F435W-F814)_{0}=1.5$ separates SSCs from GCs.}
    }
    \label{fig:color_ssc}
\end{figure*}
\begin{figure*} %
        \includegraphics[width=18cm]{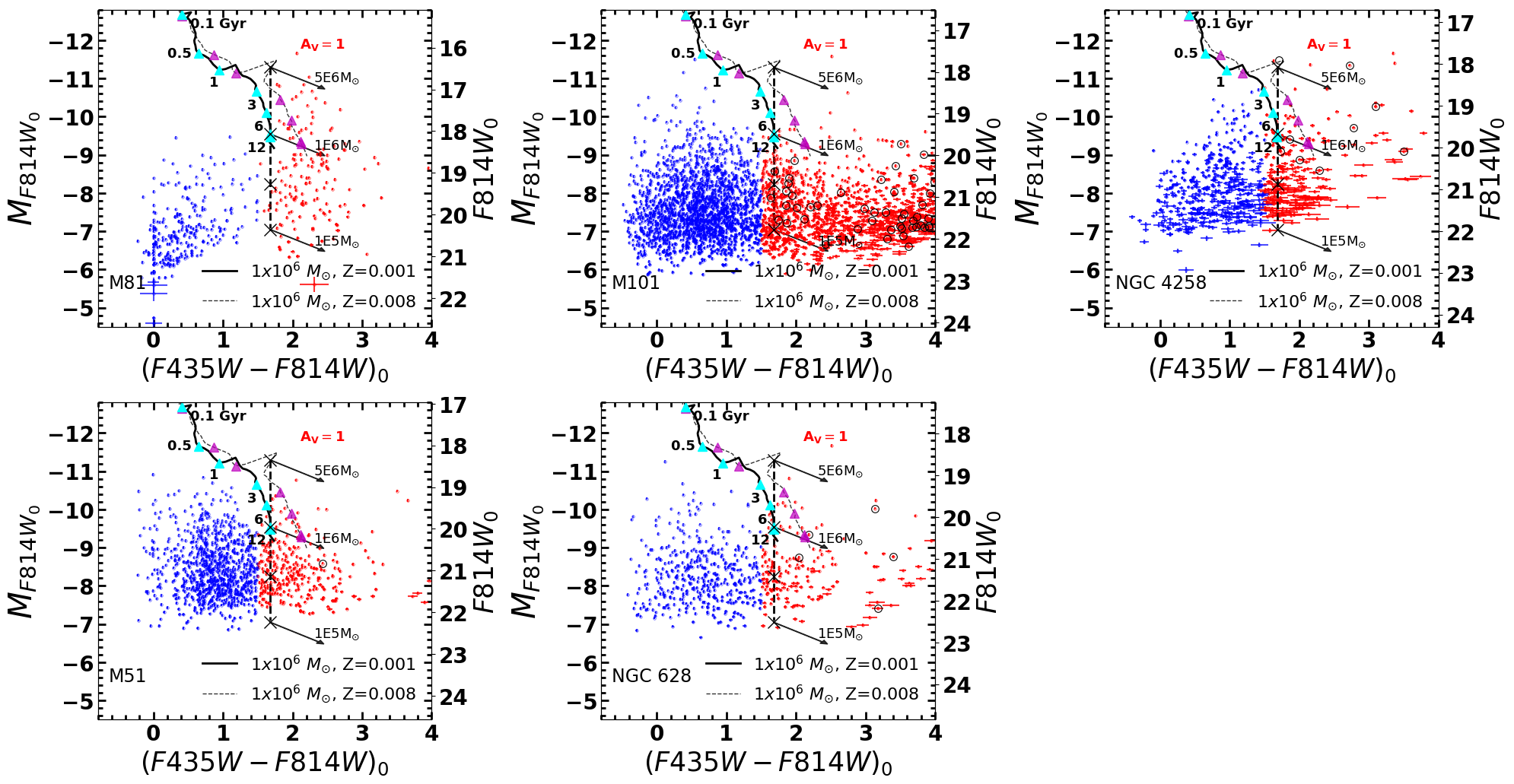}
        \caption{$M_{F814W_{0}}$ vs $(F435W-F814W)_{0}$ CMD of all cluster candidates in our sample of galaxies.
Candidates having $(F435W-F814W)_{0}>$1.5~mag are GC candidates (red dots), and
bluer objects are young disk cluster candidates (blue dots). The evolutionary
locus of the SSPs from \citet{Bruzual:2003} for two metallicities (Z=$0.001$, black solid line and Z=$0.008$, black dashed line), of mass=$1\times10^{6}$~\mso\ and Kroupa IMF,
are shown. Locations corresponding to selected ages (0.1, 0.5, 1, 3, 6 and 12~Gyr) are marked  (cyan and magenta triangles in each SSP). The chosen colour cut separates 
clusters older than 3~Gyr from the younger ones for unreddened SSPs. The reddening vectors with $A_{V}=1$~mag are shown for 3 cluster masses for an SSP of 12~Gyr age and Z=0.001, typical
values expected for GCs. The majority of the GC candidates 
are in the zone occupied by clusters of mass 1$\times10^{5}$ to 5$\times10^{6}$~\mso\ and
$A_V$ between 0 and 1~mag. Encircled red dots correspond to compact GC 
candidates ($1.07<$CI$<1.15$; see section~\ref{seccion:seleccion} for details) in our sample. 
}
    \label{fig:dcm}
\end{figure*}

\subsection{The sample of GCs}
\label{seccion:sample}

We define GCs as old (age$>$10~Gyr) metal-poor ($Z\lesssim0.001$) clusters, 
having properties similar to that for the sample of Galactic GCs. 
However, unlike the Galactic GCs, extragalactic GCs 
cannot be selected by their visual appearance even on the HST images. 
Our cluster sample contains GCs as well as
relatively young clusters, such as SSCs. 
In Table~\ref{tabla:fuentes}, we list the source detection statistics in each galaxy.
The second column contains all SExtractor-defined sources, with the column~3 containing
the number of cluster sources. Cluster samples in spiral galaxies contain 
more SSCs than GCs, and hence quantitative criteria are required to discriminate between GCs and SSCs. 
Cluster colour is the most useful discriminator for achieving this. For example,
metal-poor SSPs (Z$\leq$0.001) predict $B-I\geq$1.5~mag for populations older than $\sim$3~Gyr
\citep{Bruzual:2003}. 

\begin{table}   
\begin{center}
\setlength\tabcolsep{1.0pt}
\caption{Source detection and selection statistics.}
\begin{tabular}{l r r c r  r}      
\hline
Galaxy   &  All    & $N_{\rm SSC+GC}$   & N$_{\rm GC}$ & $N_{\rm GC}^U/N_{\rm GC}$ & $N_{\rm cont}/N_{\rm GC}$ \\
(1)      &  (2)    & (3)   &  (4) & (5)     & (6)   \\
\hline
\hline
M81     &  565438 &  433   &  158  & 0.65 & 0.20  \\
M101    & 1215533 & 3091  &  1123 & 0.14 & 0.32    \\
NGC4258 & 1360607 &  626  & 226    & 0.35 & 0.35   \\
M51     &  452747 & 1196   &  293  & 0.52 & 0.25    \\
NGC628  &  224108 &  608   &  173  & 0.41 & 0.14   \\
\hline
\end{tabular}
\label{tabla:fuentes} 
\begin{tablenotes}
    {
     {\it Notes:} (1) Galaxy. (2) All sources (stellar+non-stellar+spurious) detected by SExtractor in all the pointings over the target galaxy. (3) Those sources satisfying the criteria explained in Section~3.2 to be identified as a cluster. (4) Subset of red ($B-I>$1.5~mag) sources in column~3. (5) Fraction of total number CGs with photometry in an ultraviolet filter. (6) Fraction of contaminants (reddened SSCs) in our GC sample.}
\end{tablenotes}
\end{center} 
\end{table}

\begin{table*}
\setlength\tabcolsep{3.6pt}
\begin{scriptsize} 
	\centering
	\caption{Observational properties of the brightest three GCs in each sample galaxy.
}
\label{tabla:m101_muestra} 
\begin{tabular}{l c c c c c r c c r c c c c c c} 
\hline
ID  & RA       & DEC      &  $I$    & $(B-I)_{0}$ & $(B-V)_{0}$ & $(U-B)_{0}$ & FWHM &AREA & CI & M$_{F814W}$ & r$_{gc}$ (kpc) & FLAG \\
(1) & (2)      & (3)      & (4)   & (5)         & (6)         & (7)  &(8)  &(9)&  (10) &  (11) & (12) & (13)   \\ 
\hline
\hline             
      M81-GC1     & 148.84125& 69.110425 & 16.13$\pm$ 0.03 & 2.06$\pm$ 0.10 &  1.10$\pm$ 0.10&           1.63$\pm$0.10            &  3.62  &   -  & -  & $-$11.66 &  3.03  &   02      \\
      M81-GC2     & 148.94123& 69.050097 & 16.75$\pm$ 0.03 & 2.24$\pm$ 0.10 &  1.27$\pm$ 0.10&           2.05$\pm$0.10            &  3.45  &   -  & -  & $-$11.03 &  1.54  &   03      \\
      M81-GC3     & 149.11440& 69.019380 & 16.95$\pm$ 0.03 & 1.78$\pm$ 0.10 &  1.00$\pm$ 0.10& 1.43$\pm$0.10&  5.82  &   -  & -  & $-$10.84 &  5.87  &   02      \\
\hline                                                                 
      M101-GC1    & 210.824946& 54.318723 & 17.63$\pm$ 0.02 & 2.51$\pm$ 0.00 &  1.20$\pm$ 0.17 &   1.31$\pm$ 0.41    &  3.07  &   449   & 1.30 & $-$11.57 &  4.00  &   03      \\
      M101-GC2    & 210.884651& 54.414166 & 18.57$\pm$ 0.02 & 2.74$\pm$ 0.00 &  1.14$\pm$ 0.17 &           --            &  3.38  &   394   & 1.50 & $-$10.63 &  9.82  &   00  \\
      M101-GC3    & 210.884508& 54.369237 & 19.10$\pm$ 0.02 & 2.48$\pm$ 0.00 &  1.46$\pm$ 0.17 &           --            &  2.92  &   301   & 1.39 & $-$10.10 &  6.31  &   00      \\
 \hline                                                                
      NGC4258-GC1 & 184.729270& 47.264868 & 17.83$\pm$0.00 & 1.63$\pm$0.00 & 0.79$\pm$0.10 &  -- & 2.69 &  311 & 1.16 &  -11.57  &  5.25 & 00 \\
      NGC4258-GC2 & 184.865700& 47.210389 & 17.89$\pm$0.00 & 3.37$\pm$0.01 & 1.13$\pm$0.10 &  -- & 2.79 &  165 & 1.18 &  -11.50  & 16.78 & 00 \\
      NGC4258-GC3 & 184.745970& 47.301067 & 17.97$\pm$0.00 & 2.15$\pm$0.01 & 0.99$\pm$0.10 & 0.43$\pm$0.05 & 4.03 &  440 & 1.65 &  -11.42 &  0.69 & 02 \\
      
\hline                                                       
      M51-GC1    & 202.459584& 47.174710 & 18.90$\pm$ 0.03 & 2.05$\pm$ 0.00 &  1.07$\pm$ 0.10 &   $-$0.06$\pm$ 0.17 &  3.10  &   462   & 1.44 & $-$10.76 &  3.23  &   02      \\
      M51-GC2     & 202.507117& 47.241845 & 19.18$\pm$ 0.03 & 3.50$\pm$ 0.00 &  1.52$\pm$ 0.10 &  1.17$\pm$ 0.17 &  2.77  &   271   & 1.35 & $-$10.48 &  7.96  &   03      \\
      M51-GC3     & 202.469420& 47.253906 & 19.31$\pm$ 0.03 & 1.75$\pm$ 0.00 &  1.25$\pm$ 0.10 &           --            &  2.47  &   281   & 1.24 & $-$10.35 &  8.80  &   00      \\
\hline      
      NCG628-GC1  &  24.221429& 15.788914 & 18.28$\pm$ 0.03 & 2.51$\pm$ 0.00 &  1.08$\pm$ 0.16 &    1.82$\pm$ 0.07   &  3.33  &   471   & 1.56 & $-$11.66 &  7.84  &   03      \\
      NCG628-GC2  &  24.171023& 15.782500 & 19.13$\pm$ 0.03 & 1.90$\pm$ 0.00 &  0.68$\pm$ 0.16 &   0.28$\pm$ 0.09 &  2.75  &   243   & 1.37 & $-$10.81 &  0.52  &   02      \\
      NCG628-GC3  &  24.212973& 15.734987 & 19.71$\pm$ 0.03 & 3.15$\pm$ 0.00 &  1.25$\pm$ 0.16 &           --            &  2.62  &   183   & 1.16 & $-$10.23 &  10.48 &   00      \\
   	\hline
    \end{tabular}
            \begin{tablenotes}
                \begin{small}
                \item {\it Notes:} (1) Assigned name, which follows the convention of $GAL$-GC$n$, 
where $GAL$ stands for galaxy name and $n$ is 1 for the brightest object in the $F814W$ filter, and 
increases sequentially as the magnitude increases. 
(2,3) Right ascension, Declination, coordinates in J2000. 
(4) Magnitude in $F814W$-band and magnitude error from SExtractor and aperture correction, added in quadrature, 
(5) $(F435W-F814W)_{0}$ colour, the error is the quadrature sum of the error in each band from SExtractor. 
(6) $(F435W-F555W)_{0}$ colour, error is quadrature sum as in column~5. 
In the case of M81 the colour is $(F435W-F606W)_{0}$. 
(7) $(F336W-F435W)_{0}$ colour, the error is the quadrature sum of the error in each band.  In the case of M81 the colour is. $(u-g)_{0}$.
(8) FWHM in pixel units from SExtractor. 
(9) AREA in pixels from SExtractor. 
(10) Concentration index, defined as the difference between magnitudes 
measured in 1 and 3 pixel radius apertures. 
(11) Absolute magnitude in $F814W$-band.
(12) Galactocentric distance of the GC in kiloparsec.
(13) FLAG classification: 00 clusters without $U$-band photometry; 01 determined as a reddened young cluster from $U$-band photometry; 02 - determined as bonafide GC from $U$-band photometry; and 03 - error in U-B is likely to be larger than the indicated formal error.
                \end{small}
            \end{tablenotes}
\end{scriptsize} 
\end{table*}

In Figure~\ref{fig:color_ssc}, we plot the distribution
of $(F435W-F814W)_{0}$ colours for all clusters (black histogram) and for clusters 
brighter than $M_{F814W}=-$9~mag (red histogram) in each galaxy. The latter is 
multiplied by a factor which is indicated in the figure annotation. 
For the bright cluster sample, four of the five galaxies studied here show a 
bimodality in the distribution, the exception being M101. The colour that separates
the two distributions is nearly the same in all these galaxies, and 
corresponds to the $(F435W-F814W)_{0}$=1.5~mag bin. The full cluster sample also shows a minimum
in the distribution close to this colour in four of the galaxies.
In M101, although the evidence for bimodality is not strong for the bright cluster
sample, the total sample indicates bimodality with the saddle point again 
corresponding to the $(F435W-F814W)_{0}$=1.5~mag. The blue and red distributions correspond to the SSCs, 
and GCs, respectively. Based on this, we use $(F435W-F814W)_{0}$=1.5~mag to separate GCs and SSCs. 
Column~4 of Table~\ref{tabla:fuentes} lists the number of clusters classified as GCs using
this colour criterion.
Coordinates and photometry for each of the selected GCs are given in Table \ref{tabla:m101_muestra}.
The Table in the body of the text contains data only for three brightest objects in each galaxy. 
The entire Table is available in electronic version.

The red tail of the reddened SSC colours most likely extends beyond the colour cut of $(F435W-F814W)_{0}$=1.5~mag,
and hence colour-selected GC sample in spiral galaxies is expected to have some contamination from the reddened SSCs.
The SSC population has a long tail on the red side, which
is either due to a spread in reddening, or/and due to a spread in ages of SSCs.
In either case, the red tail of SSC distribution has very few clusters beyond $(F435W-F814W)_{0}>$1.5~mag. The estimation of contaminating objects in the GCs samples is discussed in Section \ref{seccion:contaminantes}.

In Figure \ref{fig:dcm}, we plot all candidate clusters in a colour-magnitude diagram
(CMD) formed using $M_{F814W_{0}}$ vs $(F435W-F814W)_{0}$, where SSCs and
GCs are shown by blue and red points, respectively.
The evolutionary loci of clusters for Simple Stellar Population (SSP) models 
from \citet{Bruzual:2003} at typical metallicities of SSCs (Z=0.008$\sim$1/3 solar) and 
GCs (Z=0.001) are shown. 
These models correspond to synthetic clusters of mass $=1\times10^{6}$~\mso\   
obeying the Kroupa initial mass function (IMF) between masses 0.1 and 100~\mso. 
If the GCs are as old as 12~Gyr, the range of magnitudes covered by the GCs
corresponds to mass range shown by the dashed vertical line, and the range of colours
corresponds to reddening equivalent to $A_V$=0 to $\sim$2~mag.

\begin{figure*} 
        \includegraphics[width=18cm]{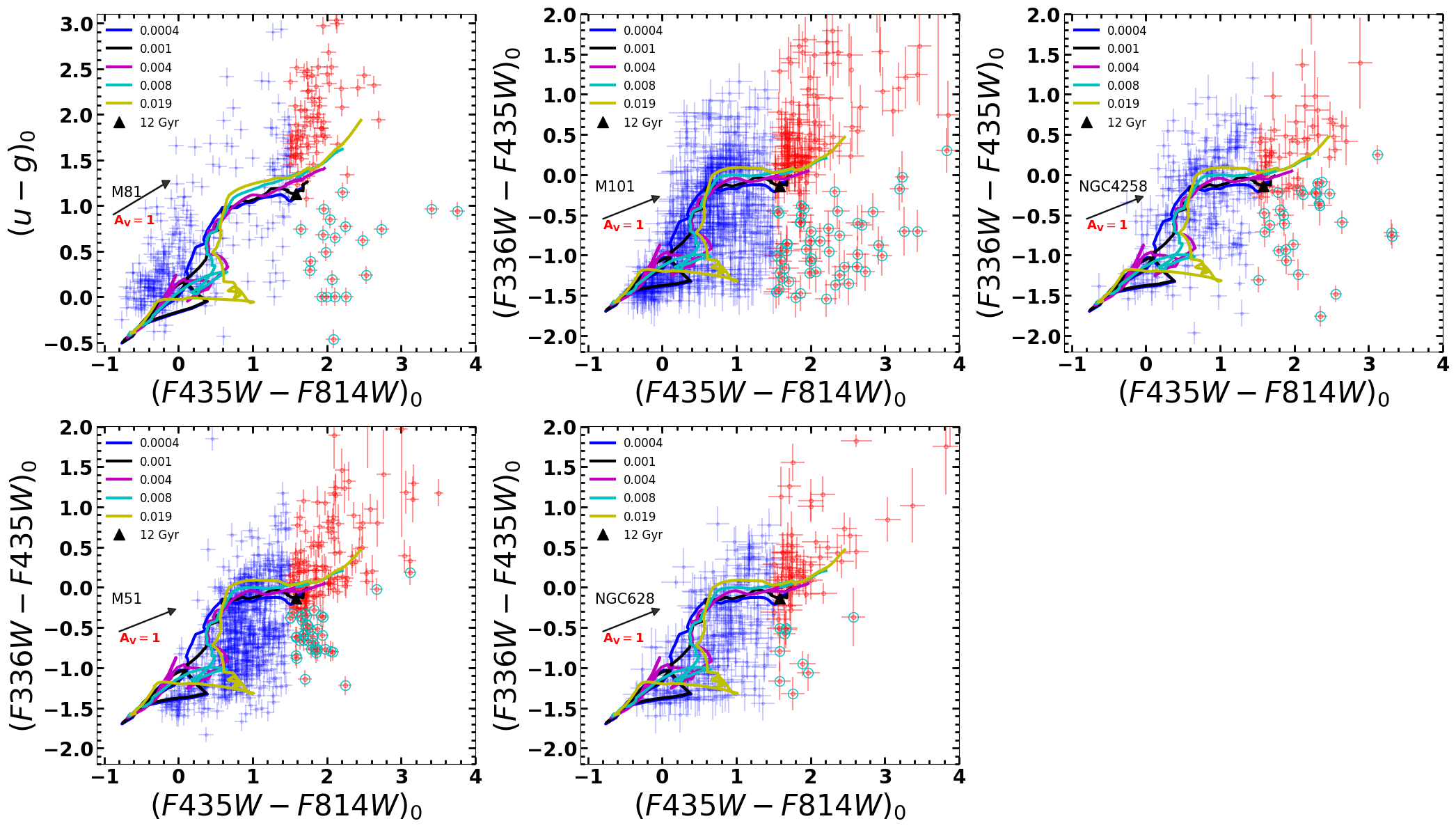}
        \caption{
        {The $U$-selected cluster candidates in $(u-g)_0$ vs $(F435W-F814W)_{0}$ diagram for M81 and $(F336W-F435W)_{0}$ vs $(F435W-F814W)_{0}$ diagram for the rest of the sample galaxies. 
Candidates having $(F435W-F814W)_{0}>$1.5~mag are GC candidates (red dots), and
bluer objects are young disk SSC candidates (blue dots). 
The evolutionary loci of SSPs from \citet{Bruzual:2003} for different metallicities
using Kroupa IMF are shown by solid curves of different colours, following the
colour notation shown in each panel. The bluest colour that a classical GC can have
(Z=0.0004 and age 12~Gyr) is marked by a solid triangle.
The reddening vector with $A_{V}=1$~mag is shown. The reddened young SSCs that occupy 
the GC colours are contaminants, which are identified by red dots surrounded by 
circles of cyan colour. 
}
}
    \label{fig:dcc}
\end{figure*}
\begin{figure*}
    \includegraphics[width=\columnwidth]{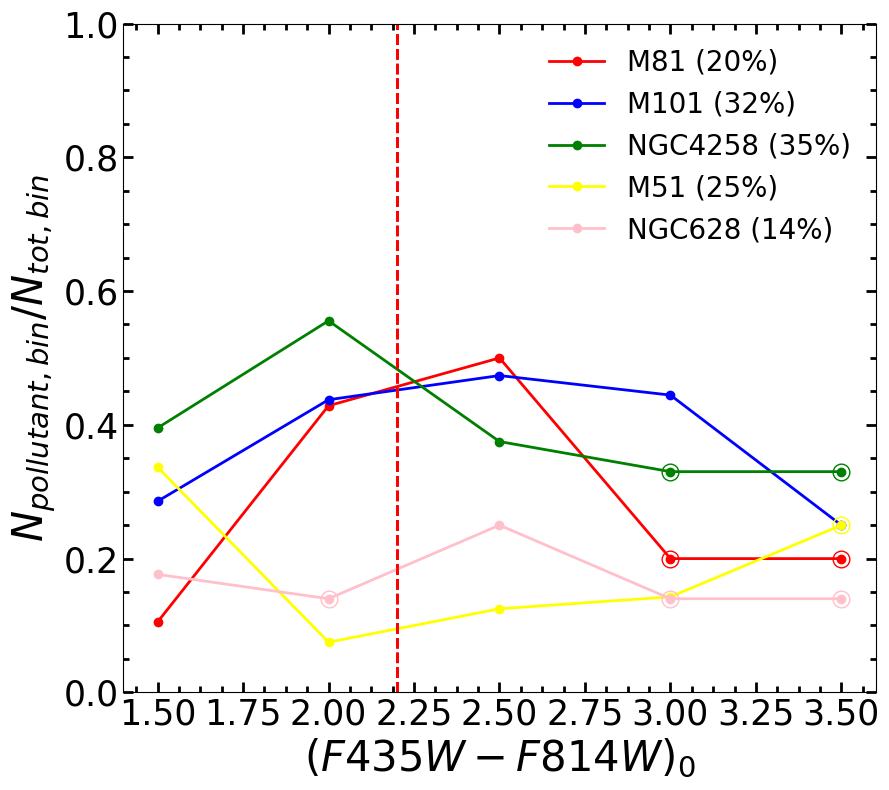}
    \includegraphics[width=\columnwidth]{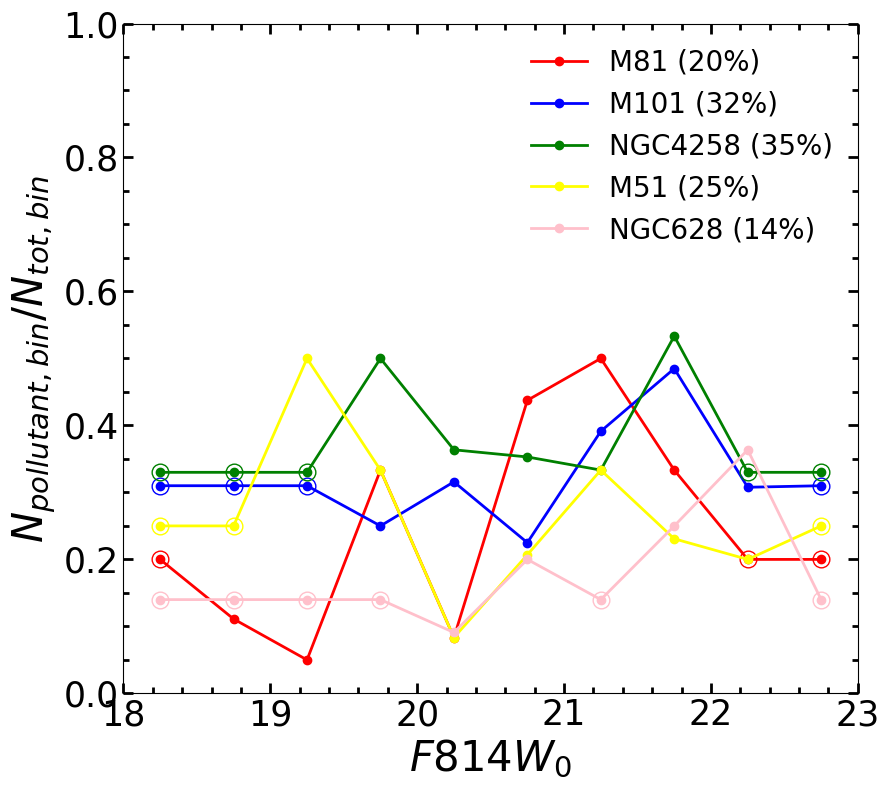}
    \caption{
    {The fraction of contaminants (reddened young SSCs) in our GC samples as a function of 
    colour (left) and magnitude (right) for our five sample galaxies. The vertical dashed-line 
    in the left figure corresponds to the reddest colours reached by SSPs for unreddened GCs.
    See text for details.
    }}
    \label{fig:cont_frac}
\end{figure*}

\subsection{Contamination of the GC sample from reddened SSCs}
\label{seccion:contaminantes}

Figure~\ref{fig:dcm} suggests that the reddened SSC colours overlap with the GC colours. 
The colour histogram (Figure~\ref{fig:color_ssc}) also suggests that the distribution of
the SSC colours (the bluer peak) most likely has a long tail on the redder side, that overlaps
with the GC colours. Thus, contamination to some degree from reddened SSCs is unavoidable 
when selecting GC samples in spiral galaxies using colour-cuts. 
We use the sub-sample of clusters with $U$-band photometry 
to estimate the fraction of contaminants in each galaxy.

Use of colour-colour diagrams involving ultraviolet and optical filters is known
to break the age-reddening degeneracy (e.g.,
\citealt{Georgiev:2006}; \citealt{Bastian:2011}; \citealt{Fedotov:2011}). 
In particular, $U-B$
colour separates clearly clusters younger (SSCs) and older (GCs) than $\sim$3~Gyr. 
Keeping this in mind, we searched the HST archives for images in the WFC3/$F336W$ filter.
All our sample galaxies have at least one pointing in this filter 
(see Table~\ref{tabla:apuntados_f336w}). 
For M101, the $F336W$ images had good astrometry. For the fields in other galaxies,
we carried out the astrometry following the same procedure as described in section 3.1. 

WFC3/$F336W$ image for M81 is available for only one pointing, as compared to the 29 pointings 
with the ACS, with only 7 GC candidates (4\%) falling in the FoV of the $F336W$ image. 
The contamination fraction obtained from such a small coverage of the FoV is not expected to be representative. 
On the other hand, Sloan Digital Sky Survey\footnote{\url{https://dr12.sdss.org/fields}} (SDSS) 
image of this galaxy obtained from multiple pointings covers the FoV of all the 29 HST/ACS pointings. 
The relative nearness of this galaxy allows the detection and photometric analysis of 65\% of clusters 
that occupy the relatively uncrowded fields. We hence used the SDSS $u-g$ colours  for estimating the 
contamination fraction in M81.
In column~5 of Table~\ref{tabla:fuentes}, we present the fraction of GC candidates with
$U$-band images, i.e. SDSS-u for M81 and HST/WFC3 $F336W$ for the rest.
We performed photometry using the {\it phot} task in {\sc iraf} using the same photometric parameters as for the HST/ACS images.

In Figure~\ref{fig:dcc}, we plot all candidate clusters with $F336W$ coverage in $(F336W-F435W)_{0}$ vs $(F435W-F814W)_{0}$  colour-colour diagram. For M81, we show the SDSS $u-g$ colours (in the AB system) in the ordinate.
The evolutionary loci of clusters in this diagram for theoretical
SSPs from \citet{Bruzual:2003} of different metallicities are shown. 
The $U-B$ colours of reddened young ($<10$~Myr) clusters are distinctly different
from that of clusters older than $\sim$3~Gyr, which allows us to break the age-reddening
degeneracy.
Thus reddened young SSCs (contaminants) would lie below the SSP locus for age$>$3~Gyr.
In other words, for a redder ($F435W-F814W_{0}>1.5$) cluster to be considered a genuine GC, its 
$U-B$ colour, after taking into account photometric errors, should correspond to
a location above the SSP locus in the figure. 
As illustrated in Figure~\ref{fig:error_estimated}, the real errors in
photometry are larger than the formal error bars, which limits the use of the
colours for a precise determination of age. 
Nevertheless, the photometric quality is good enough to
separate reddened SSCs from GCs. In the last column of Table~\ref{tabla:fuentes}, we give
the fraction of contaminants in our GC samples. The values lie between 
14--35\% in the sample galaxies. 

In Figure~\ref{fig:cont_frac}, we explore whether contaminant fraction in our GC samples 
depends on the colour and magnitude of the GCs. We carry out this analysis in bins of 0.5~mag
in colours and magnitudes. In order to avoid fluctuations caused due to small number statistics
in some of the bins, we plot only those values for which the number of contaminants in any bin
was more than the Poisson error in that bin, taken as the square root of the total number of GC 
candidates in that bin. For the rest of the bins, we plot the global values. 
In general, all galaxies have higher contaminant fraction for $(F435W-F814W)_{0}>$2.2~mag, with as
much as 50\% of the GCs of $(F435W-F814W)_{0}=$2.5~mag being contaminants (reddened young SSCs) in M101 and M81. 
None of the galaxies show any significant dependence of the contaminating fraction with magnitude.
We hence use the global values of the contaminant fraction to correct the LF obtained from 
our entire GC sample in Section~\ref{seccion:lfgc} for all galaxies.

\subsection{Comparison of our GC catalogues with those in the literature}

In four of our five sample galaxies, there exists a previous catalogue of GCs.
We reiterate that none of these catalogues are
as complete as our catalogues in terms of spatial and magnitude coverages,
as well as in the estimation of contamination fraction from reddened SSCs.
We here compare our catalogues, obtained using uniform selection criteria, 
with the catalogues from other authors, using different selection criteria.

{\it M81:} 
\citet{Nantaisfoto:2010} reported a sample of 233 GC candidates in M81,
that had made use of data from HST/ACS and SDSS.
They used a FWHM$>$3~ACS pixel (0.15$\,arcsec$) as the main discriminator. 
We find that 107 of our 154 GCs are present in the catalogue of \citet{Nantaisfoto:2010}.
Most of the remaining 49 GCs in our sample have 2.4$<$FWHM$<3$~pixel, which is the
main reason for their exclusion in \citet{Nantaisfoto:2010}.
Lower number of GCs in our sample as compared to that of \citet{Nantaisfoto:2010} 
is due to the more stringent filtering that we imposed in the selection of GCs.

{\it M101:} 
Similarly, using data from the HST/ACS, \citet{Simanton:2015} reported a sample 
of 326 star clusters in M101. The selection was made using the concentration 
index and a colour cut. Their sample includes extended objects, and hence
they used a magnitude cut of M$_{V}\leq-6.5$ to define GCs, resulting in a 
sample of 98 GC candidates. 
We find that 48 are present in their sample of GC candidates. 
Reasons for us missing 50 of their objects are that
25 of these have ELLIPTICITY$>$0.3, 19 do not meet the
AREA criterion and the rest do not meet our FWHM criterion.

{\it NGC\,4258:} 
\citet{Lomeli:2017} reported a sample of 39 GCs in NGC\,4258 using near infrared 
(NIR) data from CFHT (Canada France Hawaii Telescope) over a large FoV 
($1^\circ\times1^{\circ}$), but based on seeing-limited ($\sim$0.7\,arcsec) dataset.
The selection of the GCs was carried out using the colour-colour diagram 
$u^{*}i'K_{s}$ ($(i'-K_{s})$ vs $(u^{*}-i')$) \citep{Munoz:2014}. 
In comparison, our catalogue using HST/ACS images contains 226 GCs over a FoV of $\sim12.3^\prime\times 17.2^\prime$.
From a spectroscopic follow-up of the objects, \citet{Rosa:2019} 
concluded that the selection based on colour-colour diagram has between 10 to 30\% 
contaminants \citep[e.g.][]{Munoz:2014, Rosa:2019}.
Our FoV includes 29 objects of \citet{Lomeli:2017}, of which 23 are in our catalogue.
Out of the 6 missing objects in our sample, two are confirmed to be
GCs using spectroscopic data by \citet{Rosa:2019}, another two are dwarf galaxies, 
with the remaining one being a foreground star.
Reason for the two of their GCs to be absent in our sample is that
they do not satisfy our selection criteria.

{\it M\,51:} 
Star clusters in M51 were catalogued by \citet{Hwang:2008} using HST/ACS images.
They used the SExtractor parameters stellarity, FWHM and ellipticity for star cluster selection.
They created two subsamples, one with 2.4$<$FWHM$<$20 and the other with 2.4$<$FWHM$<$40~pixel,
resulting in a catalogue of 8400 stellar cluster candidates.
If we applied their colour discriminator ($(B-V)>$0.5 and $(V-I)>$0.8) in their catalogue
we obtained a sub-catalogue of 224 red clusters, 214 of which have FWHM$<$10~ACS ACS pixels. Only 85 of these
objects are in our catalogue of 223 GCs. 
We searched our master SExtractor output catalogue to find out the
reasons for we missing 139 of their objects,
and found that 85 of these have ELLIPTICITY$>$0.3, 15 do not meet the AREA criterion and the rest do not meet our FWHM criterion.

{\it NGC\,628:}
The last galaxy of our sample, NGC~628, has also been a target for cluster searches using HST images by \citet{Ryon:2017}. However, they reported only young and intermediate-age clusters, and no GCs.

\section{Completeness corrections}

\label{seccion:completes}
We aim to study GC luminosity functions in our sample galaxies. 
In order to do that, we need to take into account possible sources of incompleteness in our catalogues. 
Two principal sources of incompleteness in our study are: i) incompleteness in magnitude and 
ii) incompleteness in radial coverage even with multiple HST pointings. 
In this section, we describe the method we have followed to correct for these observational 
limitations and obtain a final GC luminosity function.

\subsection{Monte Carlo cluster simulations}

We generated mock clusters using the IRAF/DAOPHOT tasks $addstars$ and $mkobjects$.
A cluster is defined by an intensity profile that follows a Gaussian function 
of a given {\sc fwhm} and a total magnitude,
with {\sc fwhm} taking values of 2.0, 2.4, 3.0, 4.0, 5.0, 6.0, 7.0 and 8.0 pixels, and
magnitudes varying between 19 and 24 magnitudes at interval of 0.5~mag. For a fixed {\sc fwhm}, 1100 clusters were generated, 100 for each simulated magnitude.

\begin{figure}
    \includegraphics[width=\columnwidth]{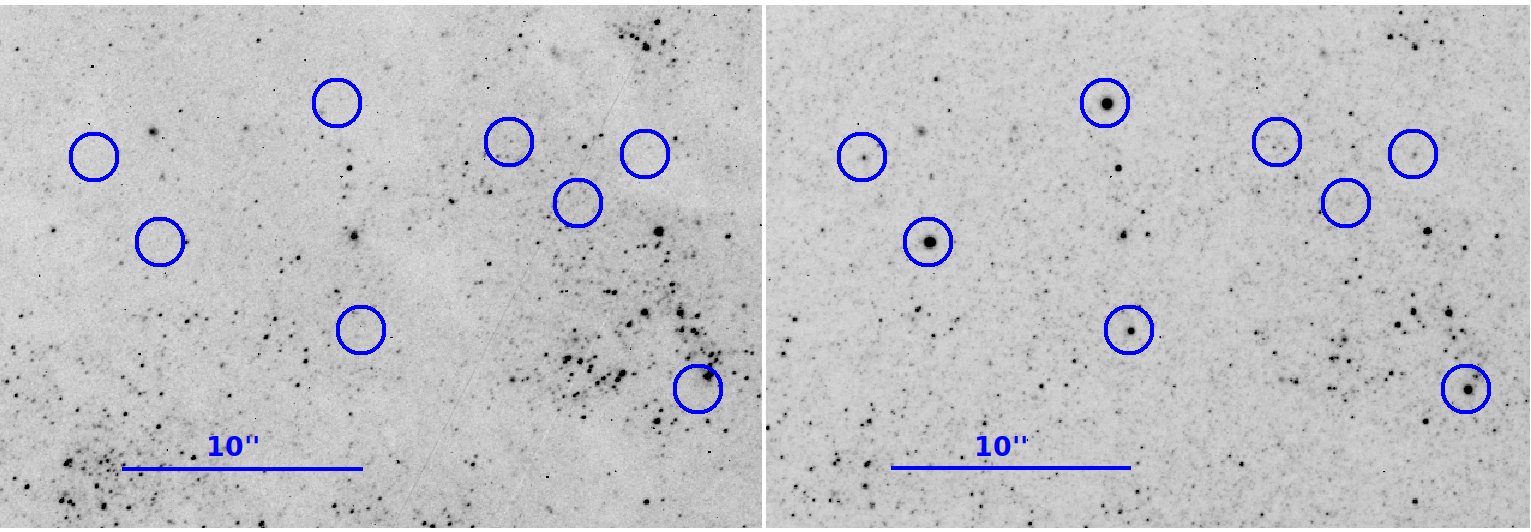}
    \caption{A randomly selected region of M101 before adding mock sources (left) and 
after adding them (right); sources have different magnitudes. All blue circular regions 
have 1\,arcsec radius. North up, east left.  }
    \label{fig:completes_fuentes_f814}
\end{figure}

Coordinates of these
sources were randomly selected and were inserted onto an observed HST image.
The faintest object that can be detected in an observation under uniform
background conditions defines the detection limit of that observation. 
In real observations, the background is often non-uniform, and each non-point
source has a different size. Further, crowding of objects in an image can lead to
non-detection of objects that are brighter than the detection limit.
Hence, for each galaxy, we chose HST images corresponding to two FoV, 
representing zones of
 (i) low background and low crowding,
and (ii) high background and high crowding. The former simulates the conditions
typical of clusters in the external parts of galaxies, whereas the latter represents the
characteristics of clusters in the bulge and spiral arms.
For M81, which has as many as 29 pointings, simulations were carried out on an additional 
frame containing the bulge and nucleus \citep[R09; see Figure 1 in][for the footprint of M81]{Mayra:2010}. 
The results of these latter simulations were used exclusively in the analysis of radial density 
distribution of clusters in the inner part of this galaxy.
The same object detection and selection criteria (\S3.2) we had used for 
real objects were applied on the mock-object added frames. 
In Figure \ref{fig:completes_fuentes_f814}, we zoom-in on a randomly selected 
region of M101, before (left) and after (right) inserting mock sources.
In the image the positions of the mock sources, which cover a range of magnitudes, 
are indicated by blue circles. 

We have also used these simulations to establish criteria to select 
cluster candidates from all sources catalogued by the SExtractor, and 
apply {\sc fwhm}-dependent aperture corrections, described in Section~\ref{seccion:aperture_correction}.
We used the results of the simulations for low background and low crowding for
this purpose.

\subsection{Completeness corrections: magnitude}

For a proper counting of a population, it is necessary to know completeness
factor as a function of object magnitude. 
We used the results of our simulation to determine the completeness factors.
This factor is expected to depend on the {\sc fwhm}.

In the top panel of Figure~\ref{fig:ci_vs_fwhm}, we plot the distribution of {\sc fwhm}, 
which peaks at {\sc fwhm}$<$3~pixel. 
Hence, we calculate the completeness factors
for mock clusters of input {\sc fwhm}=3~pixel. 
The completeness factor is defined as the ratio of the number of recovered objects 
($N_{\rm out}$) to the inserted objects ($N_{\rm in}$) at each inserted magnitude.

\begin{figure*}
\begin{center}
        \includegraphics[width=16cm]{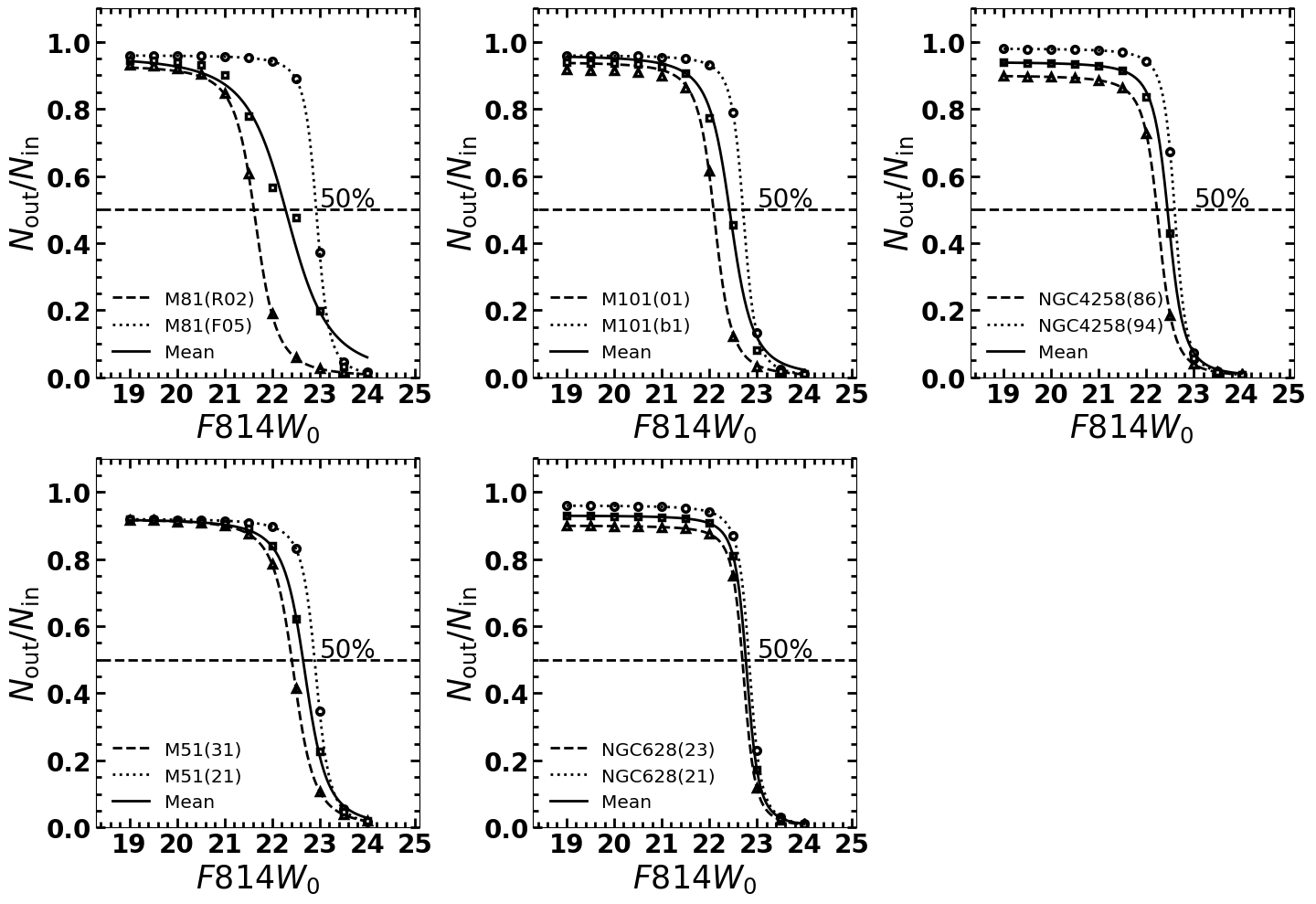}
      \caption{Completeness correction curves in the $F814W$-band for our sample galaxies
in low (dotted line passing through circles) and high (dashed line passing through triangles) 
background regions. The mean of these two curves is shown by the solid line.
The identification codes of high and low background frames used in these simulations are
shown in parenthesis.}
    \label{fig:completes_f814}
\end{center}
\end{figure*}

\begin{table}
\begin{center} 
    \setlength\tabcolsep{2.0pt}
	\caption{Magnitude at 50\% completeness in $F814W$-band, for three curves in Figure \ref{fig:completes_f814} }
	\label{tabla:completes}
	\begin{tabular}{l c c c c c c c}
\hline
Galaxy & \multicolumn{2}{|c|}{Low} & \multicolumn{2}{|c|}{High} & \multicolumn{2}{|c|}{Average} & \\
    &{$m_{\textrm{50}}$} & {$\alpha$}  & {$m_{\textrm{50}}$} & {$\alpha$} & {$m_{\textrm{50}}$}& {$\alpha$}   \\
\hline
\hline
M81      & 22.93 & 3.74$\pm$0.01  & 21.65 & 2.15$\pm$0.04 & 22.32 & 1.08$\pm$0.14  \\
M101     & 22.72 & 3.79$\pm$0.00  & 22.11 & 2.69$\pm$0.16 & 22.44 & 2.05$\pm$0.17  \\
NGC\ 4258& 22.60 & 4.02$\pm$0.01  & 22.26 & 3.02$\pm$0.00 & 22.46 & 2.96$\pm$0.13  \\
M51      & 22.92 & 3.24$\pm$0.01  & 22.45 & 2.19$\pm$0.00 & 22.69 & 2.07$\pm$0.06  \\
NGC\ 628 & 22.85 & 4.01$\pm$0.01  & 22.72 & 3.97$\pm$0.00 & 22.79 & 3.83$\pm$0.01  \\
	\hline
	\end{tabular}
           \begin{tablenotes}
                \begin{small}
                \item {\it Notes:} {$m_{\textrm{50}}$ is the magnitude at 50\% completeness and $\alpha$ is a fitting constant defined in Equation~\ref{equ:prichet}}.
                \end{small}
            \end{tablenotes}
\end{center}
\end{table}

In Figure \ref{fig:completes_f814}, we show completeness curves in the $F814W$-band
for all of our sample galaxies, for low (circles) and high (triangles) background 
frames. In order to quantitatively obtain the magnitude at which the sample is 
50\% complete, $m_{\textrm{50}}$, we fitted the points with the Pritchet function  
\citep[e.g.,][]{McLaughlin:1994, Karla:2013, Lomeli:2017} given by:
\begin{equation}
   \large{ f(m) = \frac{1}{2}\left[1 - \frac{\alpha(m-m_{\textrm{50}})}{\sqrt{1+\alpha^{2}(m-m_{\textrm{50}})^{2}}} \right] },
    \label{equ:prichet}
\end{equation}
where $\alpha$ is a fitting constant that determines curve's slope. These fitted
curves are shown by dotted and dashed lines, for low and high background frames,
respectively. 
The GCs are distributed homogeneously across galaxies, including regions 
with a high and low crowding. To make the correction for incompleteness 
of the luminosity function, we decided to use the values of the mean curve,
which is shown by the solid curve. In Table \ref{tabla:completes}, the values 
for the 50\% completeness for three curves are given for the $F814W$-band. 

\begin{figure}
  \includegraphics[width=\columnwidth]{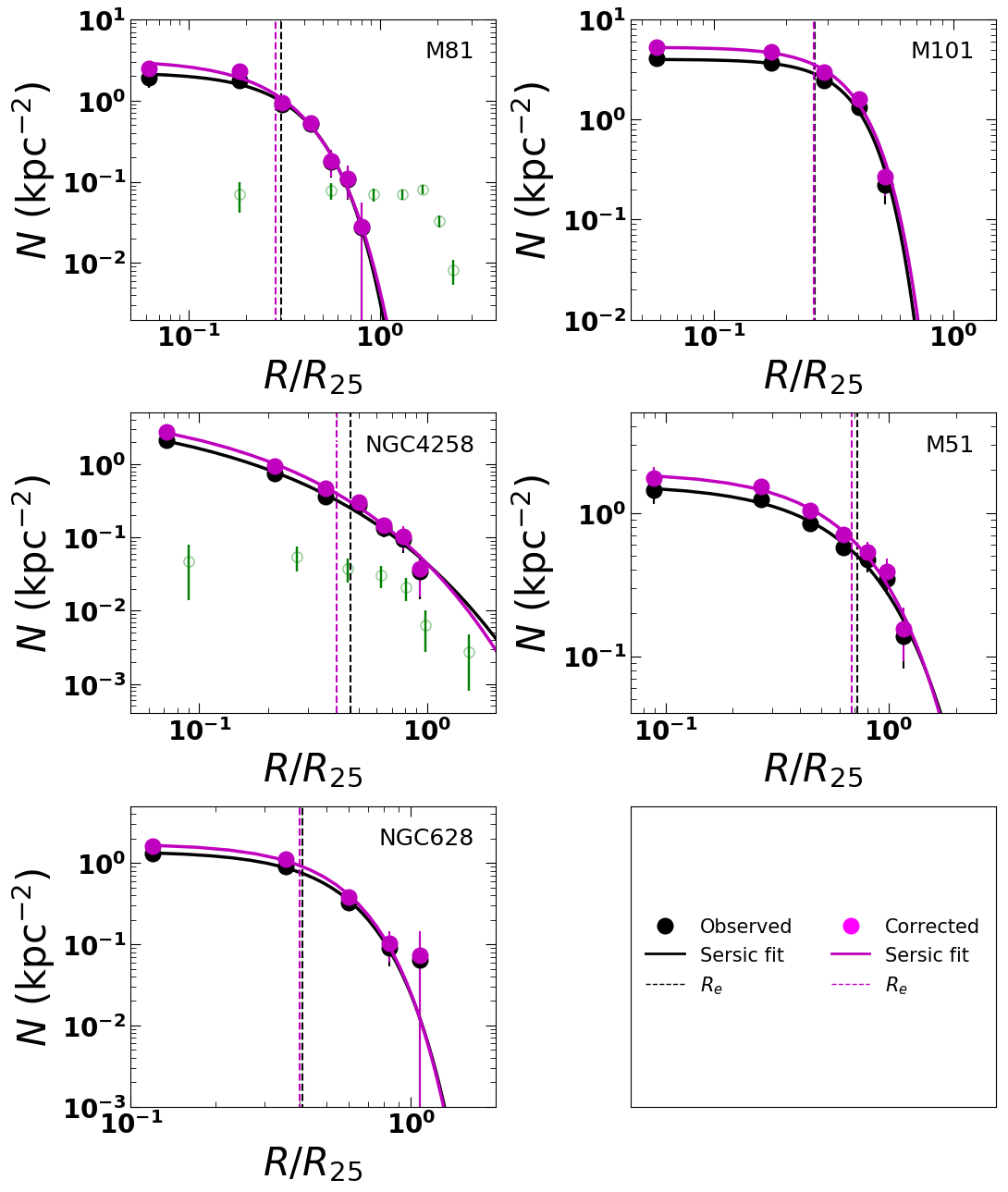}
    \caption{Radial distribution of surface number density of GCs from our data (solid circles)
are fitted with S\'ersic profile (solid line). 
Catalogued GCs beyond the spatial coverage of the HST images are shown by the empty 
circles (M81 and NGC4258, only).
Observed number densities in these two galaxies
are consistent with the extrapolation of the fitted function.
The vertical dashed lines show the best-fit values of \Rad\ before (black solid circles) and after (magenta solid circles) incompleteness correction. 
}
    \label{fig:sersic}
\end{figure}

\subsection{Completeness corrections: radial distribution}

While the images obtained from HST observations offer the best spatial resolution,
they do not cover the whole extent of galaxies even with the multiple pointings. 
Thus, outer halo GCs are often missing in catalogues selected from HST images.
The following analysis was carried out to correct for the absence of GCs beyond the galaxy's disk.

GCs are the most easily noticeable objects in the halos of galaxies.
However, their surface density in the inner parts always outnumbers that in the halos
\citep[see e.g.][]{Hargis:2014, Kartha:2014}. In fact, the radial surface density
distribution can be  well described by the S\'ersic function \citep{Sersic:1968} 
in its classical form:
    \begin{equation}
        N(R) = N_{\textrm{e}}\exp\left[-b_{n}\left(\frac{R}{R_{\textrm{e}}}\right)^{(1/n)}-1\right],
        \label{equ:sersic}
    \end{equation}
\noindent where, \Rad\ is the effective radius enclosing half the population, 
$n$ is the S\'ersic index that controls the shape of the profile and 
$b_{n}=1.992n-0.3271$. 
We show the radial distribution of surface number density in 
Figure~\ref{fig:sersic}. The radial axis is plotted in units of 
$R_{\rm 25}$.
For two of our sample galaxies (M81: \citealt{Perelmuter:1995}, 
and NGC4258: \citealt{Lomeli:2017}), 
GC searches have been carried out using the ground-based data
that complement the HST data in the outer parts, which have also been
shown (green empty circles).
The observed and completeness-corrected values are shown in solid circles of black and
magenta colours, respectively, with the fits to these data shown by solid lines
of corresponding colours. The numbers inside and outside the $R_{\rm e}$ are corrected by 
the incompleteness factors corresponding to high and low surface brightness fields, respectively. 
For M81, the inner most number is corrected by the completeness factor obtained from simulation 
on the frame that included the bulge and the nucleus.
The $R_{\rm e}$ values before and after the incompleteness corrections are shown by dashed vertical
lines of black and magenta colours, respectively. The effect of correction is to marginally
shift the $R_{\rm e}$ to lower values, which is noticeable only for NGC4258 and M51 in the figure.
In Table \ref{tabla:sersic}, we show the values for S\'ersic fit. $N_{{\rm GC},R_{\rm e}}$ and $N_{{\rm GC},R_{\rm e}}^\prime$ are the numbers 
of GCs inside \Rad\ before and after correction for contaminants.
To estimate the total number of GCs (see Section \ref{seccion:numero_total}), we use the number of GCs within \Rad, ($N_{{\rm GC},R_{\rm e}}^\prime$).

The number density obtained wide-field ground-based surveys underestimates the number of GCs inside 
$R_{\rm e}$ by more than an order of magnitude in both the galaxies with such data. 
In NGC4258, the slope of our fit in the external part agrees with the observed number densities from 
wide-field survey, with the absolute values marginally below our predictions. On the other hand, 
number densities obtained from wide-field survey in M81 remain constant over a large radial range, which 
are likely due to some selection biases in the survey. On the other hand, the wide-field survey suggests 
an excess of GCs in the external parts of this galaxy as compared to our predicted values. This excess could be due to contamination from foreground stars and unbound intergalactic GCs in the M81 group \citep[see e.g.][]{Jang:2012, Ma:2017}.

Two of the most well-studied spiral galaxies, the Milky Way and M31, have
$n=1.9$ and $R_{\textrm{e}}=4.41$~kpc and 
$n=1.6$ and $R_{\textrm{e}}=4.59$~kpc \citep{Battistini:1993}, respectively. 
In comparison, for elliptical galaxies the calculated values are larger, e.g., 
NGC 720 (E5): $n=4.16\pm1.21$, $R_{\textrm{e}}=13.7\pm2.2$~kpc; 
NGC 1023 (S0): $n=3.15\pm2.85$, $R_{\textrm{e}}=3.30\pm0.9$~kpc; 
NGC 2768 (E): $n=3.09\pm0.68$, $R_{\textrm{e}}=10.6\pm1.8$~kpc \citep{Kartha:2014}. 
The $n$ and $R_{\textrm{e}}$ for our sample galaxies are in the range of expected values in spiral galaxies.
The $n$ values are $<1$ in four of our galaxies, suggesting flatter distribution in the inner parts of our galaxies, as compared to the Milky Way and M31.

\begin{table*} 
\begin{center}
\setlength\tabcolsep{9.0pt}
\caption{Results of S\'ersic profile fits to the radial distribution of GC surface density.}
\begin{tabular}{l r r r r r r r r r r r}
\hline
Galaxy & \multicolumn{4}{|c|}{Observed} & \multicolumn{6}{|c|}{Completeness-corrected} & \\
\cline{2-5} 
\cline{8-11}
 &        $n$      &          \Rad       &  $N_{{\rm GC},R_{\rm e}}$  & $N_{{\rm GC},R_{\rm e}}^\prime$  &  &  &   $n$          &   \Rad     & $N_{{\rm GC},R_{\rm e}}$ &   $N_{{\rm GC},R_{\rm e}}^\prime$  \\
 &        &  (kpc)     &              &                  &                &   &  &    (kpc)              \\
\cline{2-5} 
\cline{8-11}
M81     & 0.57$\pm$0.07 & 4.29$\pm$0.16 & 93$\pm$11  & 74$\pm$9  & & & 0.66$\pm$0.12 & 4.04$\pm$0.23 & 109$\pm$5  & 87$\pm$4 \\
M101    & 0.33$\pm$0.05 & 7.68$\pm$0.28 & 635$\pm$24 & 432$\pm$16 & & & 0.36$\pm$0.07 & 7.61$\pm$0.32 & 802$\pm$15 & 545$\pm$10 \\
NGC4258 & 1.81$\pm$0.50 & 9.48$\pm$2.02 & 179$\pm$39 & 116$\pm$25 & & & 1.63$\pm$0.30 & 8.21$\pm$0.96 & 197$\pm$27 & 128$\pm$17 \\
M51     & 0.74$\pm$0.15 & 10.07$\pm$1.15& 250$\pm$31 & 188$\pm$23 & & & 0.71$\pm$0.12 & 9.54$\pm$0.78 & 301$\pm$20 & 226$\pm$15 \\
NGC628  & 0.48$\pm$0.08 & 6.08$\pm$0.27 & 113$\pm$7  &  97$\pm$6 & & & 0.48$\pm$0.09 & 5.95$\pm$0.29 & 149$\pm$4 & 128$\pm$3 \\
\hline
\end{tabular}
\label{tabla:sersic} 
            \begin{tablenotes}
                \begin{small}
                \item 
{{\it Notes:} $R_{\text{e}}$ is the effective radius enclosing half the population, 
$n$ is the S\'ersic index from Equation \ref{equ:sersic}. 
$N_{{\rm GC},R_{\rm e}}$ is the total number of GCs inside \Rad, whereas $N_{{\rm GC},R_{\rm e}}^\prime$ is $N_{{\rm GC},R_{\rm e}}$ after taking into account contamination from reddened young SSCs. 
}
                \end{small}
            \end{tablenotes}
\end{center} 
\end{table*}

\section{Globular cluster luminosity functions (GCLFs)}
\label{seccion:lfgc}

Having obtained a sample of GCs in 5 spiral galaxies, we now analyse the 
properties of the GC systems in these galaxies. All magnitudes and colours have 
been corrected for the foreground Galactic extinction using the $A_V$ 
values given in Table~\ref{tabla:muestra}, and the \citet{Cardelli:1989} 
reddening curve. We start our analysis with the colour distribution.

\subsection{Colour distribution}

For comparing colours of our GCs with those in the MW, we transformed all our photometry 
into the Johnson-Cousins photometric system using the transformation equation~\ref{equ:siriani_1} 
and the corresponding coefficients in Table~\ref{tabla:transformation} in Appendix~\ref{sec:transformation}, 
which were taken from \citet{Sirianni:2005}.

\begin{figure}
    \includegraphics[width=\columnwidth]{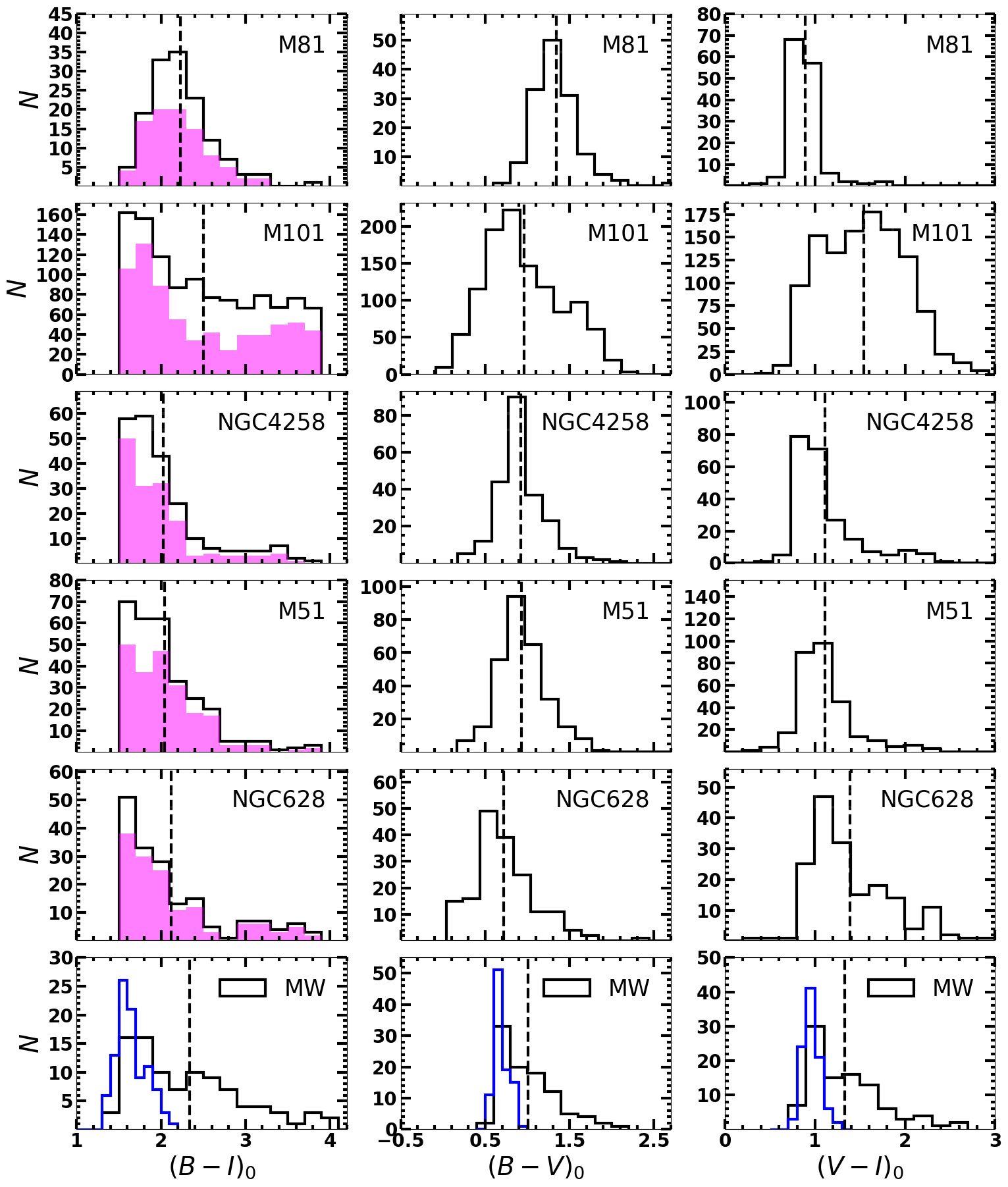}
\caption{Colour distributions of GCs of our galaxies compared to that in the MW (bottom-most panels). 
The distributions of $(B-I)_0$ colour after correcting for contaminating fractions 
are shown by magenta histograms.
The mean value for each distribution is shown by a vertical dashed line.
The colour histograms after dereddening the colours by the reddening reported for each GC for the MW GCs are shown in blue.
}
\label{fig:distribucion_color}
\end{figure}

In Figure \ref{fig:distribucion_color}, we show the distributions of 
$(B-I)_{0}$, $(B-V)_{0}$ and $(V-I)_{0}$ colours of GC systems in our sample galaxies,
where the sub-index 0 stands for Galactic reddening-corrected colours, using
the $A_V$ values in Table~\ref{tabla:muestra}. For comparison,
we also plot the colour histograms of the GCs in the Milky Way using the
data from \citet{Harris:1996}. The black and blue histograms correspond 
to the observed and dereddened colours, respectively.
In Table \ref{tabla:coloures}, we list the mean and mode of the distributions
of the three colours $(B-I)_{0}$, $(B-V)_{0}$ and $(V-I)_{0}$ for our sample
as well as for the MW sample from \citet{Harris:1996}. We use only 97 Galactic GCs 
which had integrated colours available in \citet{Harris:1996}.
For comparison, we list the colour of a SSP of 13~Gyr with Z=0.001,
the typical metallicity of old GCs, in the last row. 
The reddening-corrected colours of MW GCs agree within 0.1~mag with these SSP colours.

\begin{table}
        \setlength\tabcolsep{1.0pt}
        \centering
        \caption{The mean and mode of the colour distribution of sample galaxies and the MW.}
        \label{tabla:coloures}
        \begin{tabular}{l c c c c c c c c}
    \hline
    Galaxy    &  \multicolumn{2}{|c|}{$(B-I)_{0}$} & \multicolumn{2}{|c|}{$(B-V)_{0}$} & \multicolumn{2}{|c|}{$(V-I)_{0}$} & $A_{V}$(int)  & $A_{V}$(int)  \\
              &  Mean   & Mode          & Mean & Mode      &  Mean         & Mode   &  Mean         & Mode     \\
    (1)       & (2)     &  (3)          & (4)  & (5)       &  (6)          &  (7)  & (8) & (9) \\
    \hline
\hline 
    M81       & 2.21       & 1.81 & 1.31       & 1.21      & 0.90     & 0.77      & 0.86  & 0.32 \\ 
    M101      & 2.49 & 1.68 & 0.96  & 0.65 & 1.54& 0.98      & 1.23 & 0.15 \\ 
    NGC4258   & 2.02 & 1.64 & 0.92  & 0.81      & 1.11& 0.87 & 0.60 & 0.10 \\ 
    M51       & 2.04 & 1.71 & 0.93  & 0.87     & 1.10 & 1.03 & 0.68 & 0.19 \\ 
    NGC628    & 2.11 & 1.58 & 0.72  & 0.63     & 1.39 & 1.00 & 0.73 & 0.01 \\ 
    MW        & 2.34 & 1.70 & 1.01  & 0.72     & 1.33 & 1.01 & 1.04 & 0.17 \\ 
    \hline
    SSP       &      & 1.57 &       & 0.67     &      &  0.90 &      \\  
\hline 
        \end{tabular}
            \begin{tablenotes}
     \item {\it Notes:} (1) Galaxy name. (2--7) Mean and statistical mode of the indicated colours. (8--9) Internal visual extinction obtained from mean and mode $(B-I)_0$.
            \end{tablenotes}
\end{table}

The mode is systematically bluer than the mean in all the three colours in our sample galaxies as well as in the MW. This difference is due to a noticeable skewness in the colour distributions, with the peak (mode) occurring on the bluest bin in $(B-I)_{0}$ and $(V-I)_{0}$, in all cases except in M81.
The skewness is not due to a colour cut we have used to define GCs. In fact the choice of the cut-off colour between SSCs and GCs corresponds to the saddle point in the distribution of colours of all cluster candidates (see Figure~\ref{fig:color_ssc}), with the peak in the GC colours occurring in the first or second bin redward of the saddle point. 
We analyse the possible role of contaminants in our GC samples to the origin of skewness in colour distributions. For this, we use the fraction of contaminants as a function of colour (see left panel in Figure~\ref{fig:cont_frac}) to correct statistically the $(B-I)_{0}$ colour histograms, which is shown in Figure \ref{fig:distribucion_color} in magenta. The contamination-corrected histograms maintain the blue skewness in the distribution. 
This suggests that the skewness in colour distribution is an intrinsic property of the GC systems.

The mean and mode colours of our GC systems should correspond to SSP colours of metal-poor old clusters, if reddening is negligible. We find that the mode is only marginally redder than the colours of the metal-poor SSPs at old ages, whereas the mean is much redder. The redward shift of the mode from the expected SSP colours and the long red tail of the colour distribution, even for the MW sample, illustrate that the GC colours are often reddened by dust. The typical reddening experienced by GCs in each of our sample galaxies can be determined by comparing the mode of the colour distribution with the colour of the metal-poor old SSPs.
We made use of the $(B-I)_0$ colour to obtain the colour excess $A_V({\rm int})=((B-I)_0 - (B-I)_{\rm SSP})/(E_B-E_I)$, where $(B-I)_{\rm SSP}$ is the $(F435W-F814W)_{0}$ colour of the SSP, $E_{B}=1.33$, and $E_{I}=0.60$ are the values of the the \citet{Cardelli:1989} extinction curve at the effective wavelengths of $B$ and $I$-bands. The resulting values of $A_V$(int) using mean and mode are given respectively in columns 8 and 9 of Table~\ref{tabla:coloures}. The $A_V$(int) obtained from the mode for our galaxies range from 0.01 (NGC~628) to 0.32 mag (M81) which is comparable to the 0.17~mag for the MW. The $A_V$(int) values obtained from the mean colours vary between 0.60 (NGC4~258) to 1.23 (M101), which is also comparable to the 1.04~mag for the MW.

The difference between the mean and median $A_V$ values is $\sim$0.5--1.0 in our sample galaxies, 
which is comparable to that for the Milky Way (0.87). Thus, GC systems in spiral galaxies experience 
relatively large mean extinctions.
In fact, this is one of the reasons for the slow progress in establishing the
properties of the GC systems in spiral galaxies.
Some of the reddening could be due to dust internal to the clusters, whereas the majority
is expected due to the dust along the line of sight in the disk of the host galaxy.
The colour histogram in M101 shows a second peak at $(B-I)_0\sim$3.5~mag after correction 
for contaminants, suggesting the presence of a distinct red population. 
The analysis of LF below also suggests the presence of a population fainter than GCs. 
As we will see in Section~5.4, this population of faint red clusters corresponds to
reddened intermediate-age (1--10~Gyr) SSCs.
Under such a scenario, the basic assumption we have made in determining $A_V$, 
namely the clusters are old ($>10$~Gyr) is not applicable to all clusters in M101 and
hence the $A_V$(mean) would be an upper limit for this galaxy.

\begin{figure}
    \includegraphics[width=0.99\columnwidth]{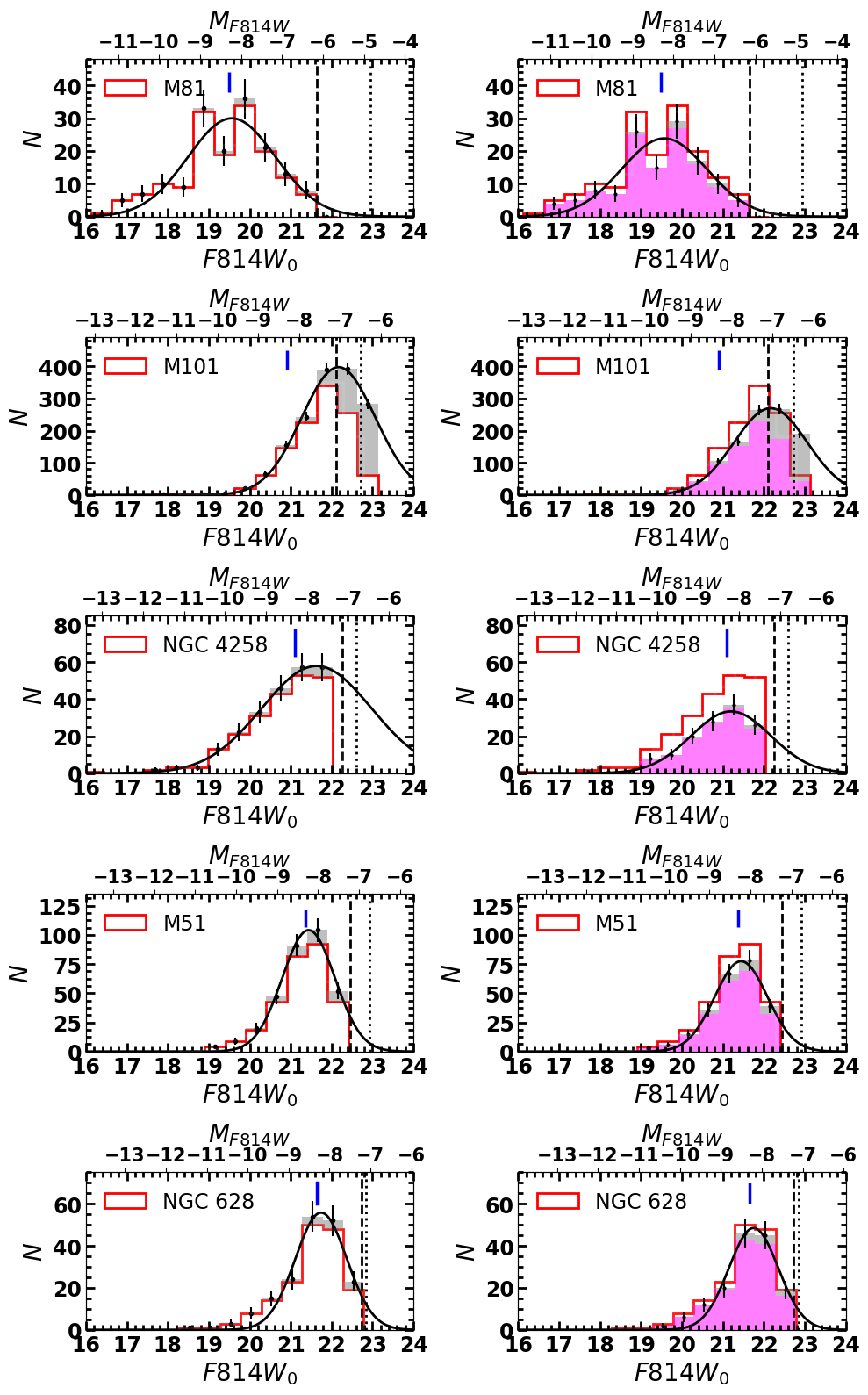}
    \caption{
{Best log-normal fits to the observed 
$F814W$-band luminosity functions of GCs in our 5 sample galaxies.
The data before (red histograms) and after (magenta-filled histograms)
correction for possible contamination from reddened young SSCs are shown in the left and right panels, respectively.
In each panel, fits were carried out to the completeness-corrected dataset (gray-filled histograms). Poisson error bars ($\sqrt{N}$) are indicated.
The vertical dot and dashed lines indicate the magnitude at which the detection is 50\% complete at low and high surface brightness parts of the galaxies, respectively.
}
    For reference, the turnover magnitude for the MW GCs is indicated in each plot by the blue vertical bar. 
    The horizontal axes in the bottom and top contain observed and absolute magnitudes, respectively.
  }
 \label{fig:gclf_full}
\end{figure}

\subsection{Luminosity function}

We chose the Galactic extinction-corrected magnitudes in the $F814W$ band 
to construct the GCLF. For this purpose, we constructed
histograms in bins of 0.5~mag over the entire range of detected magnitudes for 
each galaxy. The resulting histograms are shown in Figure~\ref{fig:gclf_full} 
with red solid lines. The magnitudes corresponding to 50\% completeness, $m_{50}$, at low and high surface brightness parts of for each galaxy are shown by dotted and dashed vertical lines, respectively. The observed numbers are
corrected for incompleteness using the mean function shown in 
Figure~\ref{fig:completes_f814}. The corrected histograms are shown by gray 
bars. As expected, the correction factor is negligible at the bright end 
($F814W\lesssim21$~mag), and gradually increases at fainter magnitudes. 

The GCLFs in the five galaxies show very similar form, increasing smoothly
up to reaching a peak value, and then again decreasing smoothly. 
In other words, all our sample galaxies show a turnover, the TO.
The TO values lie on the brighter side of $m_{50}$ in all galaxies.
The incompleteness-corrected histograms are fitted with a log-normal function given by:
\begin{equation}\label{eqn:GCLF}
dN/dM= N_{\textrm 0}e^{-( M-M_{\textrm 0})^{2}/2\sigma_{\rm M}^{2}}, 
\end{equation}
where $N_{\textrm 0}$ is a normalization factor, $M$ is the absolute 
magnitude of the fitted bin, $M_{\textrm 0}$ is the absolute magnitude of TO
and $\sigma_M$ is the dispersion.
It can be seen that the log-normal functions are good fits to the observed LFs.

\begin{table*}
\setlength\tabcolsep{4.0pt}
\caption{Number of GCs and best-fit values for a log-normal distribution.}
\label{tabla:ajsutes_gaussiana} 
\begin{tabular}{l r r  l c c c r r r r r r c c c c c c}
\hline   
Galaxy & \multicolumn{3}{|c|}{$N_{GC}$ } & \multicolumn{3}{|c|}{TO }
& $\Delta$TO & $\sigma_{M}$ &$\Delta\sigma$  & $M_{814}^{1^{st}}$ & $M_{814}^{3^{rd}}$ \\
\hline
    & Obs  & $\int$LF&  total & $F814W_{0}$ & M$_{F814W_{0}}$    & M$_{V_{0}}$  \\ 
(1)       & (2)       &  (3)    &  (4)   & (5)        & (6)             &  (7)      &  (8)           & (9)       & (10)    & (11) & (12)     \\
\hline
M81      & 126  & 128$\pm$13 & 177$\pm$15  & 19.56 $\pm$ 0.12 & $-$8.42 $\pm$ 0.13 & $-$7.52 $\pm$ 0.16 & $-$0.12 & 1.04 $\pm$ 0.17 & $-$0.11  & $-$11.66  & $-$10.84 \\
M101     & 764  & 1325$\pm$60 & 1891$\pm$70 & 22.16$\pm$0.05 & $-$7.14$\pm$0.08 & $-$6.24$\pm$0.13 &   1.16 & 0.89$\pm$0.05 & $-$0.26  & $-$11.57  & $-$10.10  \\
NGC4258  & 147  & 166$\pm$31  & 290$\pm$51 & 21.19$\pm$0.11  & $-$8.26$\pm$0.11 & $-$7.36$\pm$0.15 &    0.04 & 0.97$\pm$0.15 & $-$0.18  & $-$11.57  & $-$11.40 \\
M51      & 220  & 261$\pm$24  & 536$\pm$44 & 21.44 $\pm$ 0.07 & $-$8.34 $\pm$ 0.07 & $-$7.44 $\pm$ 0.12 & $-$0.04 & 0.64 $\pm$ 0.05 & $-$0.51  & $-$10.76    & $-$10.35 \\
NGC628   & 149  & 151$\pm$16 & 259$\pm$18   & 21.75 $\pm$ 0.05 & $-$8.21 $\pm$ 0.06 & $-$7.31 $\pm$ 0.12 &    0.09 & 0.60 $\pm$ 0.07 & $-$0.55  & $-$11.66  & $-$10.23 \\

\hline
\end{tabular}
            \begin{tablenotes}
                \begin{small}
                \item {{\it Notes:} (1) Galaxy name. (2) Number of GCs after correcting for contamination from reddened SSCs. (3) Column 2 after correcting for incompleteness in magnitude. (4) Estimated total number of GCs 
                determined using Equation~\ref{ecuacion:total_number}}. (5) Observed TO magnitude in $F814W$-band. (6) TO absolute magnitude in $F814W$-band, corrected for galactic and internal extinction. (7) TO absolute magnitude in $V$-band ($M_{V_{0}}$=$M_{F814W_{0}}$+$(V-I)_{\rm SSP}$; $(V-I)_{\rm SSP}$=0.90).  (8)  $\Delta$TO= TO(galaxy)-TO(MW) in the $V$-band, where TO(MW)=$-7.4$~mag. (9) $\sigma_{M}$, from the fit. (10) $\Delta\sigma=\sigma_M$(galaxy)$-\sigma$(MW), where $\sigma$(MW)=1.15). (11) $M_{814}^{1^{\rm st}}$, magnitude of the brightest GC. (12) $M_{814}^{3^{\rm rd}}$, magnitude of third brightest GC. 
                \end{small}
            \end{tablenotes}
\end{table*}

We obtained the TO magnitude after correcting the LF in each galaxy
for a possible contribution from the contaminants. Following the discussions in Section~3.5,
we used the magnitude-dependent contaminant fractions for NGC~4258 and 
constant contaminant fractions from the last column in Table~\ref{tabla:fuentes} for the rest.
The resulting plots for each galaxy are shown in the right panels of Figure~\ref{fig:gclf_full} 
and the results are tabulated in Table~\ref{tabla:ajsutes_gaussiana}. 
In order to determine the possible errors on the determined TO, we 
carried out 1000 Monte Carlo simulations, by adding a Gaussian noise
to each bin of the histograms. 
For this purpose, we used the square root of the number in each bin as the 
$\sigma$ of the Gaussian.
The histogram resulting from each experiment is fitted by Equation~\ref{eqn:GCLF}, 
each time finding TO and $\sigma_{\rm M}$. The rms values of these 1000 simulations are
taken as the error on TO and $\sigma_{\rm M}$.

In order to investigate the universality of the function, we converted
the apparent TO magnitudes to the absolute values using the best-available
distances for the sample galaxies (see Table \ref{tabla:muestra}), and
corrected these magnitudes for internal extinction using 
$A_{F814W}$(int)=0.58$A_V$(int), where $A_V$(int) is taken from
the last column of Table~\ref{tabla:coloures}.
The resulting values are also given in Table \ref{tabla:ajsutes_gaussiana}. 
All galaxies with the exception of M101 are consistent with ${M_I}_0=-8.21\pm0.06$~mag,
with a mean ${M_V}_0=-7.41\pm0.14$~mag. 
The TO of M101 is 1.16~mag fainter. We will discuss this case separately in next section.

How does the mean value for the galaxies compare with the values for elliptical galaxies and the MW?
TO value for the MW is $M_{\rm V}=-7.40\pm0.10$~mag (from \citealt{Jordan:2007}), which is
almost identical to the mean value of ${M_V}_0=-7.41\pm0.14$~mag obtained for
four of our galaxies.
The individual differences of the TO magnitudes for our sample galaxies from that
of the MW are given in column~8 of the Table \ref{tabla:ajsutes_gaussiana}. 
The differences are well within 1$\sigma$ value in all these four galaxies.
Hence, the GCLFs in four of our sample galaxies have the same TO as that 
in the MW. In comparison, M31 value of ${M_V}_0=-7.60\pm0.15$ obtained by
\citet{Secker:1992} and \citet{Reed:1994} is 
$\sim$0.2~mag brighter than our mean value, which is marginally larger than the
1-$\sigma$ error of our measurements.
As pointed out earlier, TO for M101 is clearly different, which could be due 
to underestimation of distance, or an intrinsically different value of TO.
This will be analysed in detail in the next sub-section.

\begin{figure}
      \includegraphics[width=0.90\columnwidth]{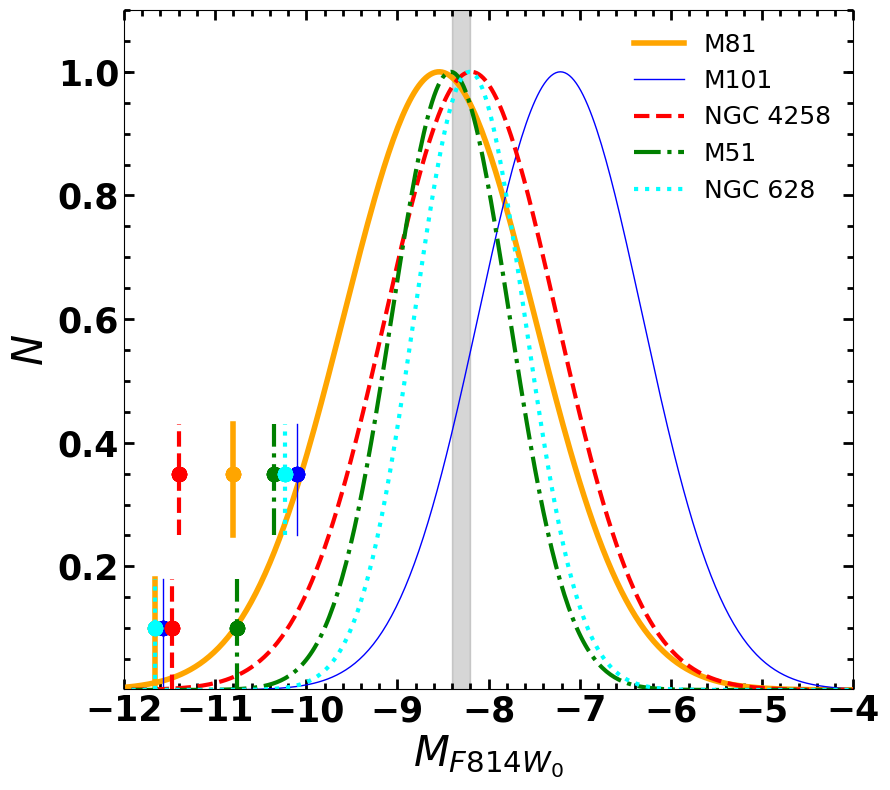}
      \caption{Comparison of the normalized F814W-band GCLFs of all our sample galaxies. 
Line types and colours used for each galaxy are indicated. The TO magnitude of the
MW GCs is marked by the gray band, with its width indicating the error on this value.
The dispersion of the TO values in 4 of the galaxies (M101 is the exception) is 
smaller than the dispersion of their brightest GCs (bottom-most set of vertical lines) or the
third brightest GCs (top set of vertical lines).
}
    \label{fig:fl_juntas}
\end{figure}

Despite the near universality of the TO values, the $\sigma_M$ of the log-normal function 
varies widely in the sample galaxies. These differences in widths are illustrated 
in Figure \ref{fig:fl_juntas}, where we plot all the log-normal fits normalized to their peak values. The TO value along with its error for the Milky Way GCs is shown as a vertical band. 
In this figure we also show the positions of the brightest GC in each of these galaxies, 
as well as the position of the third brightest GC. It can be easily seen that the dispersion 
in the brightest and the third brightest clusters in these galaxies is much higher than the 
dispersion in the TO values, suggesting that the TO values are more universal than the magnitude of the brightest, or the third brightest GC, in galaxies. The brightest GC in our sample of five galaxies belongs to M81, 
which is discussed in detail in \citet{Mayya:2013}.

\subsection{Distance errors and the universality of the TO}
\label{seccion:to_uni}

For obtaining the TO, we have used the distances tabulated in Table~\ref{tabla:muestra}. 
The tabulated distances correspond to the most recent determination of distances 
using Cepheids for M81 and M101, and the MASER distance for NGC~4258.
For M51 and NGC~628, SNII distances were used.

\begin{figure*}
    \includegraphics[height=12cm,width=18cm]{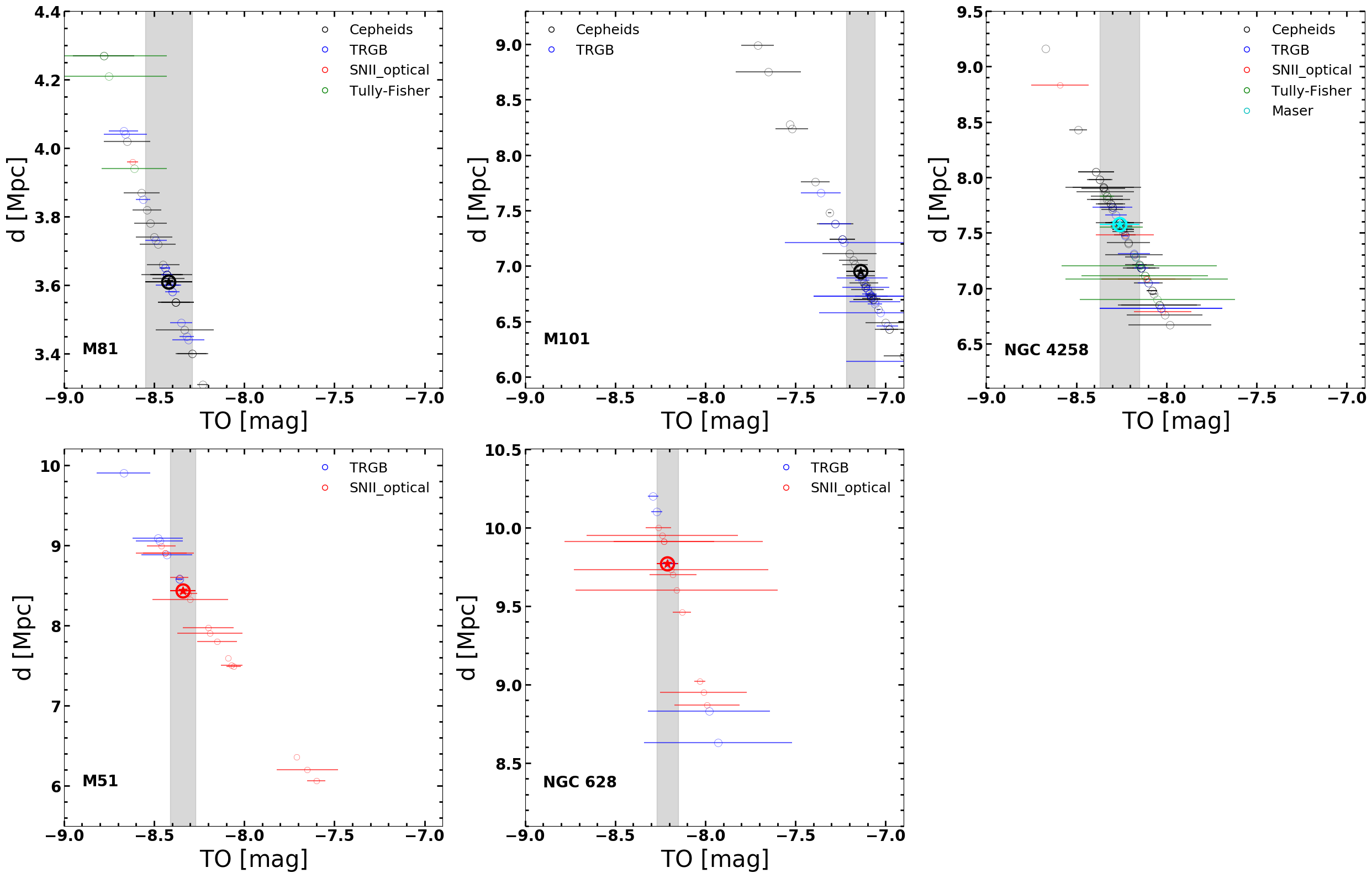}
\caption{Dependence of the measured TO (circle) on the distance uncertainties.
The measurement error on TO is shown by the width of the vertical gray band.
Symbols in each plot are colour-coded depending on the method of distance
determination.
}
\label{fig:distribucion_distancias}
\end{figure*}

Using only the distances reported after the year 2000 in 
NED\footnote{\url{https://ned.ipac.caltech.edu/}}, we found 56 independent measurements for 
M81\footnote{\url{https://ned.ipac.caltech.edu/byname?objname=m81&hconst=67.8&omegam=0.308&omegav=0.692&wmap=4&corr_z=1}}, 
79 for M101\footnote{\url{https://ned.ipac.caltech.edu/byname?objname=m101&hconst=67.8&omegam=0.308&omegav=0.692&wmap=4&corr_z=1}}, 77 for 
NGC~4258\footnote{\url{https://ned.ipac.caltech.edu/byname?objname=ngc4258&hconst=67.8&omegam=0.308&omegav=0.692&wmap=4&corr_z=1}}, 
27 for M51\footnote{\url{https://ned.ipac.caltech.edu/byname?objname=m51&hconst=67.8&omegam=0.308&omegav=0.692&wmap=4&corr_z=1}} and 
23 for NGC~628\footnote{\url{https://ned.ipac.caltech.edu/byname?objname=ngc628&hconst=67.8&omegam=0.308&omegav=0.692&wmap=4&corr_z=1}}. 
The methods of distance measurements include those using Cepheids, MASER, tip 
of the red giant branch (TRGB), SNII-optical, and Tully-Fisher. We recalculated 
TO values for each of these measured distances, which are shown plotted against 
the corresponding distances in Figure~\ref{fig:distribucion_distancias}.
The value of TO for the best-determined distance (that in Table~\ref{tabla:muestra}) 
is indicated by a thick circle, which is colour-coded according to the method 
that was used to obtain the distance. The error on this best measurement is 
highlighted by the gray band. Most of the distance measurements for the 4 galaxies 
(M81, NGC4258, M51 and NGC628) agree with the value obtained by our best distance estimate, 
after taking into consideration the errors on each of them. In M101 where we found TO 
value fainter by 1.16~mag, some of the Cepheid distances are not consistent with 
the Cepheid distance (6.95~Mpc) we have used. If we use the farthest distance reported 
for this galaxy \citep[8.99 Mpc,][]{Macri:2001}, the TO value is still 8$\sigma$ fainter 
than the universal value we have in other galaxies. 
Hence, we conclude  that the TO value in M101 is  
intrinsically fainter compared to that in other galaxies of our sample, and investigate 
the possible physical reasons for this difference.

\subsection{TO magnitude in M101 and late-type galaxies}

\begin{figure}
    \includegraphics[width=0.90\columnwidth]{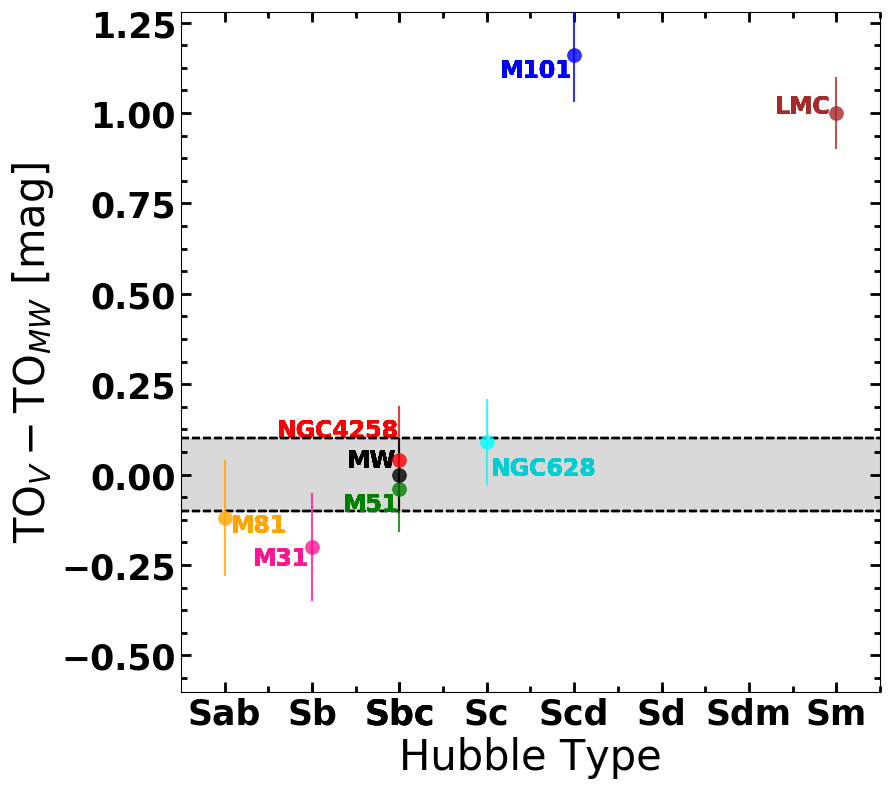}
\caption{TO$_{\rm V}-$TO$_{\rm MW}$ vs the Hubble Type of host galaxies. 
The plot includes our 5 sample galaxies,
and other spiral (M31) and irregular (LMC) galaxies where such measurements are available.
The error in the MW TO is indicated by the width of the gray band.
}
\label{fig:distribucion_TO}
\end{figure}

Among the galaxies analysed in this study, M101 stood out from the rest for having 
a fainter derived TO value, as well as for having a large population of very
red clusters ($(F435W-F814W)_{0}>$2.2~mag), including a second peak at $(F435W-F814W)_{0}>$3.5~mag in its colour 
distribution (Figure~\ref{fig:distribucion_color}). We here discuss whether these differences 
are related to its late morphological type (Scd).
In Figure~\ref{fig:distribucion_TO}, we plot the TO values with respect to
that in the MW as a function of the morphological type of the host galaxies. 
In this plot, we include the MW, M31 and the Large Magellanic Cloud (LMC). 
We estimated the TO in the LMC using data from \citet{Mackey:2003} for
 clusters with age $\geq$1~Gyr. TO values in all galaxies 
with morphological type Sc or earlier are in agreement with the value in 
the MW within the errors of measurements. The two galaxies later than Sc
deviate from this trend. Thus, there is an indication that the different TO 
value found in M101 is indeed related to its late morphological type.
In the following paragraphs, we discuss a possible scenario that naturally
explains the fainter TO in late-type galaxies.

Classical GCs are among the oldest objects in galaxies, and
their formation is related to the formation of halo and bulge \citep{Brodie:2006}. 
By definition, the prominence of these spherical components in galaxies decreases 
along the Hubble sequence.
For this very reason, most of the searches of GCs were traditionally carried out
in elliptical and early-type galaxies. Such studies have found relatively
larger population of classical GCs in ellipticals as compared to that in 
spiral galaxies (e.g., \citealt{Harris:2011}; \citealt{Harris:2013}). 
On the other hand, M101 has a weak bulge, classified as a pseudobulge by \citet{Kormendy:2010}, 
and a very poor stellar halo population \citep{Jang:2020}, as expected for its
late morphological type. These late-type systems are not expected to contain 
a large population of classical GCs. However, searches of GCs on HST images, including this study, 
find as many GCs in late-type galaxies as in early-type spirals.
It is likely that not all the inferred GCs in late-type galaxies are 
classical GCs. We analyse this issue below.

Spiral and irregular galaxies are known to contain old SSCs that were
formed in their disks and have survived the tidal shocks of their host galaxies.
For example, \citet{Rosa:2019} found that some of the objects classified
as GCs in NGC~4258, one of the galaxies analysed in our study, belong to its disk 
and share the same kinematics as the disk HI gas.
Formation and growth of galaxies through hierarchical merging, expects formation
of such clusters during every epoch of merger involving gas-rich galaxies
\citep{Mamikonyan:2017}.
The selection filters used for GC searches on HST and ground-based images do 
not distinguish these old stellar clusters from the classical GCs. 
Thus generally speaking the sample of objects observationally classified as GCs 
contains two kinds of objects, classical GCs related to the spherical component of 
galaxies, and old disk star clusters, formed {\it in-situ} or accreted onto, 
during galaxy mergers as part of the hierarchical growth of galaxies.
It is the former one that is expected to have universal TO. The fraction of old 
star clusters is expected to increase towards later Hubble types as they are
more gas-rich than the early types. Hence, the TO obtained in early-type spirals
represents that of the classical GCs, whereas in late-type galaxies it represents 
the population of old SSCs in the disk. 
This is found to be true in the LMC, where 50\% of the old cluster population
 that is used to obtain the TO has ages in the 1--10~Gyr range.

In Section~\ref{seccion:contaminantes}, we have estimated the contamination fraction from 
reddened SSCs in our sample of GCs in each galaxy, using the photometry in the $U$-band.
The technique we have used helps to distinguish reddened SSCs from GCs only for SSCs younger 
than $\sim$3~Gyr. Thus, our GC samples, in principle, contain all clusters older than $\sim$3~Gyr. 
The universality of the TO in all our galaxies earlier than Sc, suggests that these early-type spirals 
did not form significant amount of SSCs after the formation of halo and bulge,
or even if they are formed, they did not survive the gravitational shocks.

The GC population in M101 has been the subject of study by \citet{Simanton:2015}.
They found a total of 326 candidate GCs, with their luminosity function 
matching very well the LF for the Galactic GCs, but starts increasing 
rapidly for M$_{V}\geq-$6.5~mag. They identified these relatively fainter
clusters as belonging to a second population, which is statistically more 
extended than the brighter population, the so-called faint fuzzies 
(e.g.,  \citealt{Larsen:2000, Brodie:2002, Peng:2006_diffuse, Liu:2016}).
They discuss these red extended
clusters as old stellar clusters formed in the disk. Only 98 (i.e. 30\%) of their clusters 
belonged to the brighter population, the classical GCs.
Our selection criteria, which selects objects 2.4$<$FWHM$<$10 pixel, naturally 
rejects extended objects that \citet{Simanton:2015} have found, but
contains compact objects fainter than their magnitude cut of M$_{V}\geq-$6.5~mag. 
The fainter TO we have found suggests that our sample is dominated by a
population of red faint compact clusters of ages between $\sim$1--10~Gyr.
These properties correspond to relatively low-mass reddened SSCs in the disk of M101. 

\subsection{Physical parameters that control TO}
\label{seccion:to_parametros}

The TO magnitude of a GC system is controlled by four physical parameters, 
namely, the mass function (MF) at birth, age, metallicity and internal extinction.
In addition to these, a dynamical process changes the form of the mass 
function as the clusters evolve under the gravitational potential of their 
parent galaxies. 
At ages greater than a few billion years, dynamical evolution is mass-dependent,
which leads to selective destruction of low-mass clusters. 
In general, a cluster of mass $M_{\rm cl}$ loses all its mass over a 
time-scale, $t_{\rm dis}$, given by,
\begin{equation}
t_{\rm dis} = t_4\left(\frac{M_{\rm cl}}{10^4 M_\odot}\right)^\gamma,
\end{equation}
where $t_4$ is the destruction time for a cluster of $10^4 M_\odot$ under a 
specific gravitational potential, and
the power-law index $\gamma$ is found to have a value of 0.6  \citep{Boutloukos:2003}.
The $t_4$ is expected to be shorter in strong gravitational field, such as
massive early-type galaxies, and longer, in low-mass late-type galaxies.
\citet{Fall:2001} found that the net result of the dynamical evolution
is to create a MF that is log-normal, like the one found in our study,
almost independent of the initial form of the MF.
The turnover of the resulting MF is shifted to higher masses, equivalently 
to brighter magnitudes, in early-type galaxies, as compared to the late-type 
galaxies. Thus, even if the four physical parameters are the same for all GC systems,
dynamical evolution alone can introduce dispersion in the observed TOs in
different galaxies. The TO is also shifted to higher masses (brighter magnitudes)
at longer ages due to dynamical effects.  

Using our TO absolute magnitude in the $F814W$ filter, corrected for internal and 
Galactic extinction ($M_{F814W_0}^{\rm TO}$), we can determine the TO mass of the cluster 
in our sample of galaxies using an SSP of age and metallicity typical of GCs.
\begin{equation}
    { \log \mathcal{M}} = -0.4\times(M_{F814W_0}^{\rm TO}-M_{F814W}^{\rm SSP}),
\end{equation}
where $M_{F814W}^{\rm SSP}$ is the $F814W$ magnitude of an SSP at 13~Gyr for a 
total mass of 1~M$\odot$. The resulting  mass values for two of the 
lowest metallicities (Z=0.0004=1/50 solar and Z=0.001=1/20 solar) are given 
in Table~\ref{tabla:TO_physical}.
The derived masses are 3\% higher for Z=0.0004 as compared to that obtained 
using Z=0.001. In the four galaxies with almost universal TO, the mass 
corresponds to $\sim3\times10^5$~M$_\odot$, whereas it is 
lower by a factor two in M101. The disk clusters of M101 are expected to be younger 
than the uniform age of 13~Gyr we have used. If they are as young as 8~Gyr, the masses 
will be further lower by $\sim$30\%. 

\begin{table}
        \centering
        \caption{Physical parameters corresponding to TO.}
        \label{tabla:TO_physical}
        \begin{tabular}{lllllll} 
    \hline
  Galaxy  & Z & Age & $\mathcal{M}_{\rm cl}$/M$_{\odot}$ & $\sigma$  \\  
          &   & Gyr & $\log$  & $\log$                \\
    \hline
    \hline
    M81      & 0.0004 & 13.0 & 5.58 & 0.05  \\ 
             & 0.001  & 13.0 & 5.57 & 0.05 \\ 
    M101     & 0.0004 & 13.0 & 5.07 & 0.03 \\ 
             & 0.001  & 13.0 & 5.06 & 0.03 \\ 
    NGC4258  & 0.0004 & 13.0 & 5.52 & 0.04 \\ 
             & 0.001  & 13.0 & 5.51 & 0.04 \\ 
    M51      & 0.0004 & 13.0 & 5.55 & 0.02 \\ 
             & 0.001  & 13.0 & 5.54 & 0.02 \\ 
    NGC628   & 0.0004 & 13.0 & 5.50 & 0.02 \\ 
             & 0.001  & 13.0 & 5.49 & 0.02 \\ 
    MW       & 0.0004 & 13.0 & 5.53 & 0.04 \\ 
             & 0.001  & 13.0 & 5.52 & 0.04 \\

\hline
        \end{tabular}
\end{table}


\subsection{Total number of Globular Clusters and Specific frequency}
\label{seccion:numero_total}

GC searches in external galaxies using HST data are subjected to two kinds of
incompleteness: (i) magnitude incompleteness arising due to the missing 
low-luminosity GCs beyond the detection limit,
and (ii) area incompleteness due to non-coverage of outer-halo GCs because 
of the limited FoV of HST. 
The alternative strategy of searching GCs in wide-field ground-based surveys
also suffers from these two effects, with the second effect arising due
to confusion with stellar sources in the inner regions due to lack of spatial resolution.
Characterization of the LF and the radial number density distributions
with analytical functions, allows us to calculate the total number of GCs
by integrating over the entire range of these functions.
We carry out these integrations in two steps, as explained below.

First, we integrate over the fitted GCLF to obtain the number corrected for
the missing low-luminosity GCs, which are given in column~3 of Table \ref{tabla:ajsutes_gaussiana}.
We define the missing factor, $f_{\rm LF}$=\ngc($\int{LF})$/\ngc(obs). 
In order to correct for the GCs outside our FoV, we counted the
number of GCs within $R_{\rm e}$. 
By definition, total number of detectable 
GCs is twice this number. The total number of GCs, corrected for 
contamination from reddened SSCs, and incompleteness both in luminosity and volume is:

\begin{equation}
   N_{{\rm GC,TOT}} = f_{LF}\times 2\times N_{{\rm GC},R_{\rm e}}^\prime, 
   \label{ecuacion:total_number}    
\end{equation}
where $N_{{\rm GC},R_{\rm e}}^\prime$ is listed in the last column in Table~\ref{tabla:sersic}, which is the number of GCs within $R_{\rm e}$ corrected for the contamination from reddened young SSCs.
In column~4 of Table~\ref{tabla:ajsutes_gaussiana}, we tabulate the total number of GCs in each sample galaxy. 

\begin{figure}
\begin{center}
    \includegraphics[width=0.95\columnwidth]{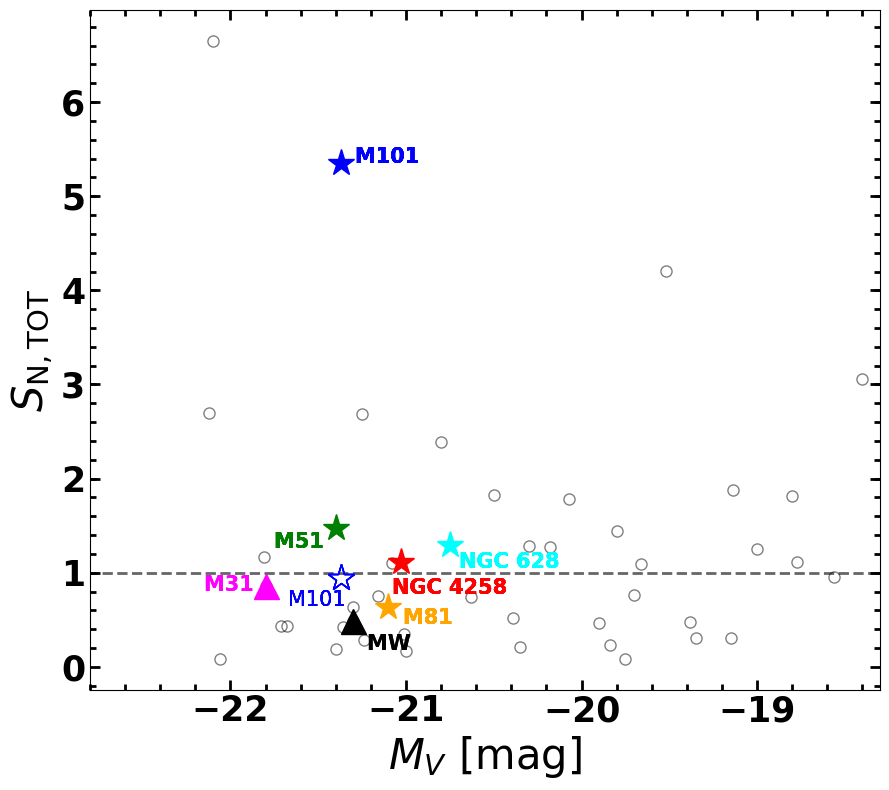}
    \caption{GC specific frequency plotted against the absolute magnitude of the host galaxy
for our sample of 5 galaxies (filled star symbol), the MW and M31 (triangles). 
{The empty star symbol for M101 corresponds to the value obtained assuming a universal TO
for this galaxy (see text for details).}
Dashed horizontal line indicates average values for spiral galaxies 
from the compilation of \citet{Harris:2013}, which are shown by empty circles.
}
    \label{fig:frecuencia}
\end{center}
\end{figure}

The specific frequency, \sn, defined as the number of GCs normalized to the total galaxy magnitude of $M_{V}=-15$~mag, is related to the $N_{{\rm GC,TOT}}$ by the relation
\sn\ $={N}_{\rm GC,TOT}\times10^{0.4({M}_{V}+15)}$ \citep{Harris:1981}.
We calculated \sn\ for each galaxy using the $M_{V}$ tabulated in Table \ref{tabla:muestra}.
In Table \ref{tabla:frecuencia} we show \sn\ and $S_{\textrm{N,TOT} }$  calculated 
with $N_{\text{GC,OBS}}$ and $N_{\text{GC,TOT}}$, respectively.
In Figure \ref{fig:frecuencia}, we show the $S_{\textrm{N,TOT} }$ vs $M_{V}$.
We also show corresponding values for the MW and M31 \citep{Peacock:2010}.
The value of \sn\ for M101 is expected to be an upper limit, given that 
our GC sample for this galaxy includes a large population of intermediate-age SSCs.
We hence made an alternative estimation of \sn, assuming that the true GC population 
in this galaxy also has the universal TO. We hence counted the number of GCs brighter 
than $M_{\rm V}=-7.40$~mag, and doubled that number to calculate $N_{{\rm GC,TOT}}$. T
he empty symbol for M101 in the figure corresponds to this revised \sn. The mean value for our sample,
including the revised value of M101 is $\sim$1.10$\pm$0.24, which occupies
a much smaller range as
compared to the values for spiral galaxies from the compilation of 
\citet{Harris:2013} (open circles). The large spread in the latter sample
is expected as they are based on studies using very different detection and analysis
techniques from 
\citet{Fischer:1990}, \citet{Harris:1996}, \citet{Kissler-Patig:1999},
\citet{Harris:2000}, \citet{Goudfrooij:2003}, \citet{Olsen:2004}, \citet{Rhode:2004}, \citet{Rhode:2007}, \citet{DiNino:2009}, \citet{Mora:2009}, \citet{Nantais:2010}, \citet{Peacock:2010}, \citet{Rhode:2010}, \citet{Adamo:2011} and \citet{Young:2012}.

\begin{table}
\begin{center}
\caption{Specific frequency.}
\begin{tabular}{l c c c c c }
\hline
Galaxy      &   \sn       & $S_{\textrm{N,TOT} }$ \\
\hline
\hline
M81             & 0.46$\pm$0.10      & 0.64$\pm$0.10\\
M101$^\dagger$  & 2.16$\pm$0.46 & 5.35$\pm$0.82\\
NGC~4258        & 0.57$\pm$0.26 & 1.12$\pm$0.30\\
M51             & 0.61$\pm$0.17      & 1.48$\pm$0.22\\
NGC~628         & 0.75$\pm$0.17      & 1.30$\pm$0.24\\
\hline
\end{tabular}
\label{tabla:frecuencia} 
\begin{tablenotes}
    $^\dagger$Values estimated assuming the universality of TO applies to M101 are 0.47$\pm$0.27 and 0.94$\pm$0.33.
\end{tablenotes}
\end{center} 
\end{table}

\section{Conclusions}

We used the HST images of five nearby spiral galaxies for which there 
are data taken with multiple pointings with ACS to enable a search
of GCs in them. Images in $F435W$, $F555W$ and $F814W$ filters were used.
We used SExtractor and a set of cluster-defining filters to define
a sample of GCs in each of the analysed galaxies. Detection of simulated
clusters was carried out to obtain the incompleteness as a function of magnitude. 
We used the $U$-band photometry for a subset of sample GCs using the SDSS $u$-band images for M81 and HST/WFC3 $F336W$ images for the rest to evaluate the contamination of our GC samples by reddened SSCs. The contaminating fraction was found to be between 0.14--0.35.
We used the dataset to construct luminosity function of GCs in spiral galaxies.

We find that the LF of GCs in all the five galaxies analysed is log-normal in 
nature, with the turn-over (TO) happening on the brighter side of the 50\% incompleteness
limit in all galaxies. This has enabled us to determine precise TO magnitudes
for our sample of galaxies. We also determined typical internal extinction towards the population
of GCs in our sample galaxies by comparing the mode of the observed colours with 
that from population synthesis models for a metal-poor (metallicity = 1/20$^{th}$
solar) population of 13~Gyr. We used the best determined distances for
the sample galaxies to convert the extinction corrected $V$-magnitude corresponding
to the turn-over into the absolute magnitude $M_V{_0}$(TO) and obtained errors on 
this by carrying out Monte Carlo simulations.
The mean of the $M_V{_0}$(TO) in 4 of our galaxies is $-$7.41$\pm$0.14, which is in excellent agreement
with the values determined
for GC population in the Milky Way of $M_V{_0}$(TO)$_{\rm MW}=-7.40\pm$0.10.
In the fifth galaxy M101, $M_V{_0}$(TO) is 1.16~mag fainter 
than that for the MW.
We propose that this difference in $M_V{_0}$(TO) arises due to morphological 
differences, with spiral galaxies
of the Hubble types Sc or earlier having a universal $M_V{_0}$(TO), whereas the 
Hubble types later than Sc have fainter $M_V{_0}$(TO). The universality of
$M_V{_0}$(TO) in early-type spirals is due to the classical GCs dominating
the GC population, whereas in late-type spirals GC population is often
dominated by old disk clusters, which are in general less massive, and
hence fainter than the classical GCs, but otherwise share the same
observational properties as the classical GCs.
The universal TO value corresponds to a stellar mass of $\sim3\times10^5$~M$_\odot$,
where the corresponding value is a factor of 2 lower for M101.

We used the dataset of GCs to address three other topics of interest, namely
the form of the radial density distribution, distribution of colours, 
and the specific number density. 
The S\'ersic function with index between $n$=0.36--1.63 is a good fit
for the radial density distribution of GCs in our sample of galaxies. 
The index value suggests exponential form, rather than the much steeper n=3--4 form found in
elliptical galaxies \citep{Kartha:2014}. 
We carried out a statistical analysis of $(B-I)_{0}$, $(B-V)_{0}$ and $(V-I)_{0}$ colour distributions
for our sample of 5 galaxies. We found that the correction factor for contaminating reddened SSCs
is a function of GC colour. Taking into account the errors in this correction factor, we find 
no compelling evidence for a bimodal distribution.
We have calculated \ngc, the total number of GCs by making corrections for 
undetected faint sources, and those outside of our FoV, by extrapolating the GCLF to fainter magnitude and 
radial surface density profiles to outside our FoV, respectively.
The specific number density \sn$\sim$1.10$\pm$0.24 for our sample of spiral galaxies.

\section*{Acknowledgements}

We are grateful to two anonymous referees for their valuable suggestions on the original manuscript,
especially for suggesting us to carryout corrections for
possible contamination of our GC samples from reddened young clusters. LLN thanks CONACyT for granting PhD research fellowship that enabled him to carry out the work presented here. We also thank CONACyT for the research grants CB-A1-S-25070 (YDM), and CB-A1-S-22784 (DRG). 

This work has made use of data from the European Space Agency (ESA) mission
{\it Gaia} (\url{https://www.cosmos.esa.int/gaia}), processed by the {\it Gaia}
Data Processing and Analysis Consortium (DPAC,
\url{https://www.cosmos.esa.int/web/gaia/dpac/consortium}). Funding for the DPAC
has been provided by national institutions, in particular the institutions
participating in the {\it Gaia} Multilateral Agreement.

\section*{Data availability}

The data underlying this article are available in the article and in its online supplementary material.




\bibliographystyle{mnras}
\bibliography{example} 




\appendix

\section{Comparison with the HLA catalogs}
\label{apen:comparasion}

We started our work using the SExtractor catalogs for cluster selection,
available at the Hubble Legacy Archive (HLA).
With our method of selection, we expect that an object
has at least AREA=50 pixels to be considered as candidate for a star cluster. 
We found that AREA values reported in HLA catalogs: 02, a2 and 11 of M101 are 
very small, with very few objects having $AREA>50$~pixel. 
On other fields of the same galaxy, there were sources having $AREA>50$~pixel. 
In order to understand the absence of sources having $AREA>50$~pixel, 
we downloaded the images and ran the SExtractor using the same 
input file as that used by the HLA. 

\begin{figure}
	\includegraphics[width=\columnwidth]{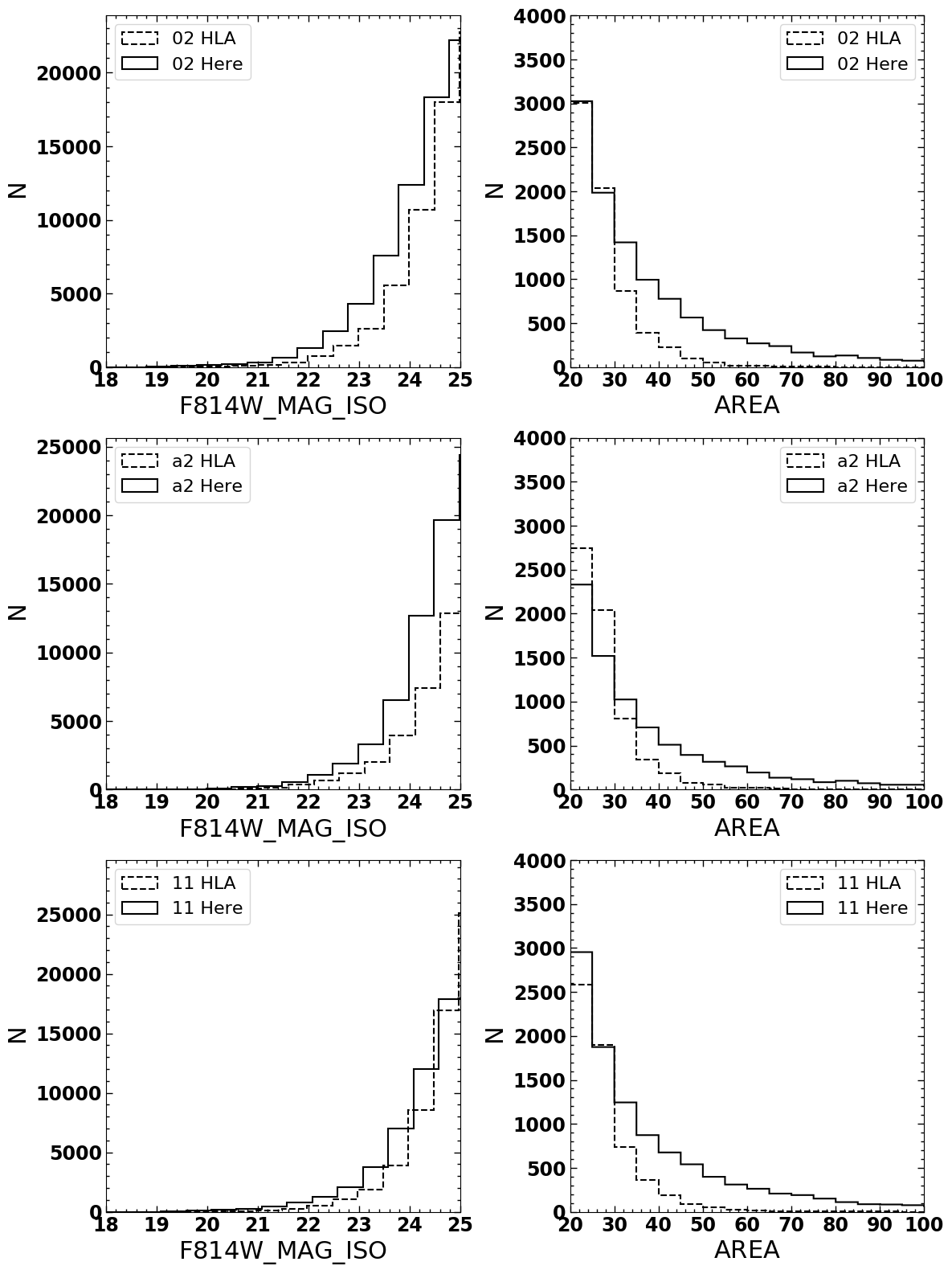}
        \caption{Frames covering M101 with incorrect values of AREA and MAG\_ISO values
in the source catalogues of HLA. The frame numbers for set are indicated. 
Dashed histograms show the values from the HLA catalogues, whereas
the solid histograms show values from our catalogue, using the same 
SExtractor input file as for the HLA catalogue.
}
    \label{fig:hla_catalog}
\end{figure}

In top Figure \ref{fig:hla_catalog}, we show a comparison between 
{\text MAG\_ISO} (top row) and AREA (bottom row) of HLA catalogs 
and SExtractor catalogues obtained by us in $F814W$-band.
The histograms of both the quantities compare very well. However, HLA 
{\text MAG\_ISO} is systematically fainter and the AREA systematically smaller
than the values we obtained, and hence the observed differences 
in fields 02, a2, 11 for M101 can be directly understood in terms of
an underestimation of background in the HLA catalogues.

The distribution of the SExtractor parameters available at HLA coincided
with that obtained from our own runs of SExtractor for all other frames 
where we made a comparison. However, in order to ensure uniform analysis 
of all sample galaxies, we downloaded the fits files and generated our own 
catalogues for each frame. The discussion presented here serves as a
cautionary note while planning to use the HLA catalogues for quantitative 
scientific studies.

\section{Transformation to Johnson-Cousins system}\label{sec:transformation}

Majority of the photometric data for GCs in the MW and external galaxies
are reported in the standard Johnson-Cousins $UBVRI$ system \citep{Bessell:1990}.
Hence, in order to compare the results obtained from our study to that 
obtained in other galaxies, it is necessary to transform our $F435W$, $F555W$ and $F814W$
magnitudes and colours into the standard system. The corresponding transformation
equations are discussed in detail by \citet{Sirianni:2005}. The transformation 
from a HST system (SOURCE), to a TARGET system (T), in our case, the Johnson-Cousins 
$UBVRI$ system, is given by:
\begin{equation}
   {\text {TMAG}} = {\text {SMAG}} + c0 + c1\times {\text {TCOL}} + c2\times {\text {TCOL}}^{2},
    \label{equ:siriani_1}
\end{equation}
where {\text{SMAG}} and {\text{TMAG}} are the magnitudes in the source and target 
systems, respectively,  {\text {TCOL}} is the colour in the target system in the
chosen bands, $c0$ is the zeropoint, which is taken from Table~\ref{tabla:apuntados} 
for the respective frames, $c1$ and $c2$ are the first and second order colour coefficients,
whose values for the colour ranges of interest in this study are given in 
Table~\ref{tabla:transformation}. Since TCOLs are in the TARGET system, 
their values are assumed to be the same as in the HST system to start with, 
which are updated iteratively, until the value of TCOLS converges.

\begin{table}
        \setlength\tabcolsep{1.9pt}
        \centering
        \caption{Transformation coefficients between HST and Johnson-Cousins systems.}
        \label{tabla:transformation}
        \begin{tabular}{l c c r r c c} 
\hline
        HST & $UBVRI$ & TCOL & c1  & c2  & Range\\
       \hline
       \hline
       F435W  & $B$ & $B-I$ & $0.022\pm0.006$ & $-0.038\pm0.011$ & $<$1.0 \\
       F435W  & $B$ & $B-I$ & $0.008\pm0.006$ & $-0.005\pm0.001$ & $>$1.0 \\
       F555W  & $V$ & $B-V$ & $-0.083\pm0.032$ &  $0.020\pm0.115$ & $<$0.2 \\
       F555W  & $V$ & $B-V$ & $-0.087\pm0.007$ &  $0.004\pm0.002$ & $>$0.2 \\
       F814W  & $I$ & $B-I$ & $0.013\pm0.006$ & $-0.028\pm0.112$ & $<$0.3 \\
       F814W  & $I$ & $B-I$ & $-0.010\pm0.010$ & $-0.006\pm0.002$ & $>$0.3 \\
\hline 
       F606W  & $V$ & $B-V$ & $0.170\pm0.005$ & $0.061\pm0.002$ & ... \\
\hline
        \end{tabular}
\end{table}

\bsp	
\label{lastpage}
\end{document}